\documentclass[a4paper,10pt]{article}
\usepackage{amsmath,amsthm,amssymb,exscale}
\usepackage{amsfonts}
\usepackage{mathtools}
\mathtoolsset{showonlyrefs}
\usepackage{epsfig,color}
\usepackage{graphicx,psfrag}
\usepackage{enumitem}
\usepackage{bbm}
\usepackage[german, english]{babel}

\DeclareMathOperator{\Tr}{Tr}
\DeclareMathOperator{\sgn}{sgn}
\DeclareMathOperator{\ad}{ad}
\DeclareMathOperator{\Ad}{Ad}
\DeclareMathOperator{\Fix}{Fix}

\newtheoremstyle{Theorem}{3pt}{3pt}{}{}{\itshape}{---}{.5em}{}
\theoremstyle{plain}
\newtheorem{thm}{Theorem}[section]
\newtheorem{lem}{Lemma}[section]
\newtheorem{cor}{Corollary}[section]
\newtheorem{rem}{Remark}[section]

\title{Equivalence of Domains for Hyperbolic
Hubbard-Stratonovich Transformations}

\author{J.\ M\"uller-Hill and M.R.\ Zirnbauer}

\begin{document}
\selectlanguage{english}

\maketitle

\begin{abstract}
  We settle a long standing issue concerning the traditional
  derivation of non-compact non-linear sigma models in the theory of
  disordered electron systems: the hyperbolic Hubbard-Stratonovich
  (HS) transformation of Pruisken-Sch\"afer type. Only recently the
  validity of such transformations was proved in the case of
  $U(p,q)$ (non-compact unitary) and $O(p,q)$ (non-compact orthogonal)
  symmetry. In this article we give a proof for general non-compact
  symmetry groups. Moreover, we show that the Pruisken-Sch\"afer type
  transformations are related to other variants of the HS
  transformation by deformation of the domain of integration. In
  particular we clarify the origin of surprising sign factors which
  were recently discovered in the case of orthogonal symmetry.
\end{abstract}

\section{Introduction}

Non-compact non-linear sigma models are an important and extensively
used tool in the study of disordered electron systems. The relevant
formalism was pioneered by Wegner \cite{weg79}, Sch\"afer \& Wegner
\cite{schweg80}, and Pruisken \& Sch\"afer \cite{prusch82}. Efetov
\cite{efe83} improved the formalism by developing the supersymmetry
method to derive non-linear sigma models. Many applications of the
supersymmetry method can be found in the textbook by Efetov
\cite{efe97}.

There exist different ways to derive non-linear sigma models from
microscopic models; for an introduction see \cite{zirn06}. One step in
the traditional approach uses a Hubbard-Stratonovich transformation,
i.e., a transformation of the form
\begin{align} \label{motiv_identity}
    c_0 \, e^{-\Tr A^2} = \int_{\mathrm{D}} e^{-\Tr Q^2 - 2i \Tr QA}
    |dQ| ,
\end{align}
where $c_0 \in \mathbb{C}$ and the domain of integration D is left
unspecified for now. $|dQ|$ denotes Lebesgue measure of a normed
vector space.

For the case of compact symmetries the transformation is just a
trivial Gaussian integral. To give an indication of the difficulty
which arises in the case of a non-compact symmetry (also known as the
boson-boson sector of Efetov's supersymmetry formalism) let us
briefly discuss the example of orthogonal symmetry $O(p,q)$. There,
$A$ is given by $A_{ij} = \sum_{a=1}^N \Phi_{a,i} \Phi_{a,j}s_{jj}$
with $s = \mathrm{Diag}(\mathbbm{1}_p, -\mathbbm{1}_q)$ and
$\Phi_{a,j} \in \mathbb{R}$. The $\Phi_{a,j}$ represent the
microscopic degrees of freedom. Using equation \eqref{motiv_identity}
and integrating out $\Phi$ gives a description in terms of the
effective degrees of freedom $Q$. The task is to find a domain of
integration D for which identity \eqref{motiv_identity} holds and the
term $\exp(-2i\Tr QA)$ stays bounded. The latter condition is imposed
in order for Fubini's theorem to apply, as further execution of the
Wegner-Efetov formalism calls for the $\Phi$ and $Q$ integrals to be
interchanged.  Note that the real matrices $A$ obey the symmetry
relation $A=sA^ts$. A naive choice of integration domain D keeping
the term $\exp(-2i\Tr QA)$ bounded would be the domain of all real
matrices satisfying $Q = sQ^ts$. Unfortunately, this choice of D is
not a valid choice in the context of the integral
\eqref{motiv_identity} as it renders the quadratic form $\Tr Q^2 =
\Tr QsQ^ts$ of indefinite sign.

Sch\"afer and Wegner (SW) \cite{schweg80} suggested a domain and
showed that it solves the difficulty. Yet, a different domain was
proposed in later work by Pruisken and Sch\"afer (PS)
\cite{prusch82}. Until recently the mathematical status of identity
\eqref{motiv_identity} for the PS domain was unclear. The main
obstacle in proving \eqref{motiv_identity} for the PS domain is the
existence of a boundary. This precludes an easy proof by completing
the square and shifting the contour (as is possible for the standard
Gauss integral and for the SW domain). Nevertheless, the PS domain
was used in most applications worked out by the mesoscopic and
disordered physics community; an early and influential paper of this
kind is \cite{VWZ85}. Most likely, the reason is that it is easier to
do calculations with, as it is invariant under the full symmetry
group of the domain of matrices $A$.

Recently Fyodorov, Wei and Zirnbauer in a series of papers
\cite{fyod05,fyod07,zirn08} proved the PS variant of the HS
transformation for the special cases of unitary and orthogonal
symmetry. In this article we extend the results to more general
symmetry groups. Moreover, our proof clarifies the relation between
the PS transformation, the SW transformation and the standard
Gaussian integrals. It is shown that the different integrals can be
transformed into each other by deforming the domain of integration
without changing the value of the integral.

Here is a guide to reading: In section \ref{Results} we define the
setting and state our main result in the form of a theorem. In
addition, we give two corollaries which relate more directly to
previous results. In section \ref{Examples} we apply our results to
three different symmetry classes. In particular, previous results
concerning the cases of unitary and orthogonal symmetry are
reproduced. The proof of the theorem is contained in section
\ref{Proof}, which is divided into three subsections. For the
convenience of the reader each subsection is preceded by a short
introduction of notation, essential structures, and a lemma
containing the results of the pertinent part of the proof. The last
subsection of section \ref{Proof} deals with the two corollaries.

\section{Statement of result}
\label{Results}

All constructions take place in $\mathfrak{gl}(n,\mathbb{C})$, the
Lie algebra of complex $n\times n$ matrices. [Please be advised
however that the following results also apply to the case where
$\mathfrak{gl}(n, \mathbb{C})$ is replaced by a complex reductive Lie
subalgebra of $\mathfrak{gl}(n, \mathbb{C})$.] Let $s\in
\mathfrak{gl}(n, \mathbb{C})$ be hermitian with the property $s^2 =
\mathbbm{1}$. This matrix $s$ gives rise to two involutions
$\theta(X) = sXs^{-1}$ and $\gamma(X) = - s X^{\dagger}s^{-1}$ on
$\mathfrak{gl}(n,\mathbb{C})$. `Involution' here means an involutive
Lie algebra automorphism. For greater generality we allow for further
involutions $\tau_i$ to be present on $\mathfrak{gl}(n,\mathbb{C})$.
Two requirements have to be fulfilled: Firstly, all involutions have
to commute with each other and secondly, $s$ has to be in the plus or
minus eigenspace of each $\tau_i$, i.e. $s = \eta_i \tau_i(s)$ with
$\eta_i \in \{\pm 1\}$.

The fixed point set of $\gamma$ and the $\tau_i$'s is the real Lie algebra
\begin{align*}
  \mathfrak{g}=\{X \in \mathfrak{gl}(n,\mathbb{C}) \mid X= \gamma(X)
  \; \mathrm{and}\; \forall i \; : \; X = \tau_i(X) \} .
\end{align*}
We also introduce the real vector space
\begin{align*}
    \mathcal{Q}= \{Q \in \mathfrak{gl}(n,\mathbb{C}) \mid Q = - \gamma(Q)
    \; \mathrm{and}\; \forall i \; : \; \,Q = \eta_i \tau_i(Q) \} ,
\end{align*}
which is an $\mathbb{R}$-module for the adjoint (or commutator) action
by $\mathfrak{g}$.

Due to $(\theta \circ \gamma)(X) = - X^{\dagger}$, the decompositions
of $\mathfrak{g}$ and $\mathcal{Q}$ into the plus and minus one
eigenspaces of $\theta$ are decompositions into hermitian and
antihermitian parts. We write these decompositions as $\mathfrak{g} =
\mathfrak{k} \oplus \mathfrak{p}$ and $\mathcal{Q} = \mathcal{Q}_+
\oplus \mathcal{Q}_-$, where $\mathfrak{k}$ and $\mathcal{Q}_+$ are
in the plus one eigenspace and $\mathfrak{p}$ and $\mathcal{Q}_-$ are
in the minus one eigenspace. $\mathfrak{k}$ and $\mathcal{Q}_-$
consist of antihermitian matrices whereas $\mathfrak{p}$ and
$\mathcal{Q}_+$ consist of hermitian matrices. The commutation
relations among all these spaces,
\begin{align*}
  \begin{matrix*}[l] [\mathfrak{k}, \mathfrak{k}] \subset \mathfrak{k}
    , & [\mathfrak{k}, \mathfrak{p}] \subset \mathfrak{p} ,&
    [\mathfrak{p}, \mathfrak{p}] \subset \mathfrak{k} , &
    [\mathcal{Q}_+ \,, \mathcal{Q}_-] \subset \mathfrak{p} , \\
    [\mathcal{Q}_{\pm} \, , \mathcal{Q}_{\pm}] \subset \mathfrak{k} ,&
    [\mathfrak{k}, \mathcal{Q}_{\pm}] \subset \mathcal{Q}_{\pm} \, , &
    [\mathfrak{p}, \mathcal{Q}_{\pm}] \subset \mathcal{Q}_{\mp} \, ,
 \end{matrix*}
\end{align*}
imply that $\mathfrak{g} \oplus \mathcal{Q}$ is a Lie algebra. (This
Lie subalgebra of $\mathfrak{gl}(n,\mathbb{C})$ could have served as
the starting point of our setting.) By the definition of $\mathcal
{Q}$ the matrix $As$ is hermitian for all $A \in \mathcal{Q}$. Note
that $s \in \mathcal{Q}_+$. To preclude any pathologies that might
otherwise occur, we demand that the Lie group $\exp(\mathfrak{k})$ be
closed.

The parametrization of the Pruisken-Sch\"afer domain is given by
\begin{align}
  PS: \; \mathfrak{p} \oplus \mathcal{Q}_+ & \rightarrow \mathcal{Q}
  ,\notag \\ (Y,X) &\mapsto e^Y X e^{-Y} . \label{ps_pk_param}
\end{align}
The standard domain for a Gaussian integral is called `Euclidean' in
the following. It is parametrized by
\begin{align*}
  Euclid: \; &\mathcal{Q}_- \oplus \mathcal{Q}_+ \rightarrow
  \mathcal{Q}^{\mathbb{C}} , \\
  &(\tilde{Y},X) \mapsto X + i \tilde{Y} ,
\end{align*}
where $\mathcal{Q}^{\mathbb{C}}= \mathcal{Q} \oplus
i\mathcal{Q}$. Finally, the parametrization of the one-parameter
family of Sch\"afer-Wegner domains is given by
\begin{align}
  SW: \; \mathfrak{p} \oplus \mathcal{Q}_+ &\rightarrow
  \mathcal{Q}^{\mathbb{C}} , \notag \\ (Y,X) &\mapsto X - i b e^Y s
  e^{-Y} ,\label{SW_parametrisation}
\end{align}
where $b$ is any positive real number.

The following statement relies on making a choice of orientation for
the PS domain. (Note that no such choice is made for D in
\eqref{motiv_identity}.) Once and for all we now fix an orientation
for each of the vector spaces $\mathcal{Q}_+$, $\mathfrak{p}$, and
$\mathcal{Q}_-$. By viewing $PS$, $Euclid$, and $SW$ as
orientation-preserving maps, we then have orientations on the
corresponding domains of integration.

\begin{thm} \label{thm_2_1} Let $A \in \mathcal{Q}$ in the setting
  above. If $As > 0$ one has
\begin{align*}
  \lim_{\epsilon \rightarrow 0}\int_{PS} e^{-\Tr(Q^2) - 2 i \Tr(Q A)}
  \chi_{\epsilon}(Q) dQ = c \, e^{-\Tr(A^2)} .
\end{align*}
Here, $\chi_{\epsilon}(Q)= \exp(\frac{\epsilon}{4}\Tr [ Q-\theta(Q)
]^2) \leq 1$ is a regulating function ($\epsilon > 0$) and $dQ$
denotes a constant volume form (i.e.\ a constant differential form of
top degree) on $\mathcal{Q}$. The normalization constant $c\in
\mathbb{C} \setminus \{0\}$ does not depend on $A$.
\end{thm}

The main idea of the proof is to show that the PS domain can be
extended by a nulldomain (of holomorphic continuation of $dQ$) and
then deformed into a Euclidean domain without changing the value of
the integral. Appendix \ref{app_SW_Euclid} shows that one can also
deform the SW domain into this Euclidean domain. Thus the PS and SW
domains are deformations of the same Euclidean domain.

Now we formulate two corollaries. For that purpose let $\mathfrak{h}$
be a maximal Abelian subalgebra $\mathfrak{h} \subset \mathcal{Q}_+
\subset \mathfrak{gl}(n,\mathbb{C})$. We require that $s \in
\mathfrak{h}$. Let $\Sigma_+(\mathfrak{k}\oplus \mathcal{Q}_+,
\mathfrak{h})$ denote a set of positive roots of the adjoint action
of $\mathfrak{h}$ on $\mathfrak{k}\oplus \mathcal{Q}_+$. Similarly
$\Sigma_+(\mathfrak{p} \oplus \mathcal{Q}_-, \mathfrak{h})$ denotes a
set of positive roots of the adjoint action of $\mathfrak{h}$ on
$\mathfrak{p}\oplus \mathcal{Q}_-$. The multiplicity of a root
$\alpha$ is denoted by $d_{\alpha}$.

The following corollary is the analogue of corollary 1 in \cite{zirn08}.
\begin{cor} \label{cor_2_1} Let $|dg|$ denote Haar measure of the closed
analytic subgroup $G \subset GL(n,\mathbb{C})$ with Lie algebra
$\mathfrak{g}$. We then have
\begin{align*}
    \lim_{\epsilon \rightarrow 0} \int_{\mathfrak{h}} \left(\int_G
    e^{-2i\Tr(g\lambda g^{-1}A)} \chi_{\epsilon}(g\lambda g^{-1})
    |dg| \right) e^{-\Tr \lambda^2} J'(\lambda) |d\lambda| = \tilde{c}
    \, e^{-\Tr( A^2)} ,
\end{align*}
where $|d\lambda|$ denotes Lebesgue measure on the vector space
$\mathfrak{h}$ and
\begin{align*}
  J'(\lambda) = \prod_{\alpha \in \Sigma_+(\mathfrak{k}\oplus
    \mathcal{Q}_+,\mathfrak{h})} |\alpha(\lambda)|^{d_{\alpha}}
    \prod_{\alpha \in \Sigma_+(\mathfrak{p}\oplus
    \mathcal{Q}_-, \mathfrak{h})} \alpha(\lambda)^{d_{\alpha}} .
\end{align*}
The constant $\tilde{c}\in \mathbb{C}\setminus \{0\}$ does not depend
on $A$.
\end{cor}

\begin{rem} It is particularly noteworthy that for odd multiplicities
  $d_{\alpha}$ of roots $\alpha \in \Sigma_+(\mathfrak{p}\oplus
  \mathcal{Q}_-,\mathfrak{h})$ the `Jacobian' $J'(\lambda)$ is not
  positive but has alternating sign.
\end{rem}

The following corollary is the analogue of theorem 1 in \cite{zirn08}:
\begin{cor}\label{cor_2_2} If the parametrization $PS$ is nearly
  everywhere injective and regular, then
\begin{align*}
  \lim_{\epsilon \rightarrow 0} \int_{\mathrm{Im} PS} e^{-\Tr(Q^2) - 2
    i \Tr(Q A)} \chi_{\epsilon}(Q) \sgn (J'(\lambda)) |dQ| = \tilde{c}'
    e^{-\Tr( A^2)} .
\end{align*}
Here $\mathrm{Im} PS = PS$ denotes the non-oriented image of $PS$. The
mapping from $\mathrm{Im} PS$ to $\mathfrak{h}$ sending $Q$ to
$\lambda$ is well defined up to a set of measure zero. $|dQ|$ denotes
Lebesgue measure on $\mathcal{Q}$, and $\tilde{c}'\in \mathbb{C}
\setminus \{0\}$ is a constant which does not depend on $A$.
\end{cor}

\begin{rem}
While we believe that the assumptions on $PS$ in corollary
\ref{cor_2_2} follow from the general setting, we have not been able
to find a proof thereof.
\end{rem}

\section{Examples} \label{Examples}

First we reproduce the examples of unitary and orthogonal symmetry.
For this we calculate $J'$ and apply corollary \ref{cor_2_1}.

\subsection{$U(p,q)$ symmetry}

This case has been handled by Fyodorov \cite{fyod05} using different
methods. To apply the general theorem \eqref{thm_2_1} we work in the
complex Lie algebra $\mathfrak{gl}(p+q,\mathbb{C})$ and define $s=
\mathrm{Diag}(\mathbbm{1}_p, -\mathbbm{1}_q)$. No additional
involutions $\tau_i$ are needed. We have
\begin{align*}
    \mathfrak{k}\oplus \mathcal{Q}_+ = \{x\in \mathfrak{gl}(n,\mathbb{C})
    \mid X = sXs \} , \quad \mathfrak{p} \oplus \mathcal{Q}_- = \{x \in
    \mathfrak{gl}(n,\mathbb{C}) \mid X = -sXs \} .
\end{align*}
The maximal Abelian subalgebra $\mathfrak{h} \subset \mathcal{Q}_+$
is spanned by the real diagonal matrices. Let $\lambda :=
\mathrm{Diag} (\lambda_1, \dots, \lambda_{p + q}) \in \mathfrak{h}$
be such a matrix. The roots $\Sigma_+(\mathfrak{gl}(p+q) ,
\mathfrak{h})$ are given by $f_i - f_j$ where $i < j$ and
$f_i(\lambda) = \lambda_{i} $. For $i\leq p < j \leq p+q$ the roots
$f_i - f_j$ are elements of $\Sigma_+(\mathfrak{p}\oplus
\mathcal{Q}_-, \mathfrak{h})$, otherwise they are elements of
$\Sigma_+(\mathfrak{k}\oplus \mathcal{Q}_+,\mathfrak{h})$.

The root space corresponding to $f_i-f_j$ is $\mathbb{C}E_{ij}$ where
$E_{ij}$ is the matrix with unity in the $ij$ position and zero
elsewhere. Thus every root space has complex dimension one, or real
dimension two. Hence
\begin{align*}
  J'(\lambda) = \prod_{i<j} |\lambda_{i} - \lambda_{j}|^2 .
\end{align*}
With this expression for $J'(\lambda)$ the formula of corollary
\ref{cor_2_1} agrees with that of Fyodorov \cite{fyod05}.

\subsection{$O(p,q)$ symmetry}

This case has been dealt with by Fyodorov, Wei and Zirnbauer
\cite{zirn08}. In addition to the involutions of the unitary setting
we need an involution $\tau_1(X) = -sX^ts$ and $\eta_1 = -1$. The
additional presence of this involution requires all matrices to be
real. In consequence, all root spaces are now one dimensional, and
they give rise to non-trivial signs:
\begin{align*}
  J'(\lambda) &= \prod_{\alpha \in \Sigma_+(\mathfrak{p}\oplus
    \mathcal{Q}_-, \mathfrak{h})} \alpha(\lambda) \prod_{\alpha \in
    \Sigma_+(\mathfrak{k}\oplus \mathcal{Q}_+,\mathfrak{h})}
  |\alpha(\lambda)| \\
  &= \prod_{i \leq p < j \leq p+q} (\lambda_i- \lambda_j) \prod_{i<j
    \leq p, p < i < j \leq p+q} |\lambda_{i}- \lambda_{j}| \\
  & = \prod_{i<j} |\lambda_{i} - \lambda_{j}| \prod_{i=1}^{p}
  \prod_{j=p+1}^{p+q} \sgn(\lambda_{i} - \lambda_{j}) ,
\end{align*}
which is precisely corollary 1 in \cite{zirn08}.

\subsection{$Sp(2p,2q)$ symmetry}

Now we consider the case of symplectic symmetry which arises for
random matrix ensembles of class $A$II in the language of
\cite{zirn96}. Let $\sigma^i$ $(i = 1, 2, 3)$ denote the three Pauli
matrices and let $\sigma^0 = \mathbbm{1}_2$.  Introducing $\sigma_p^i
= \mathbbm{1}_p \otimes \sigma^i$, we choose $s =\mathrm{Diag}
(\sigma_p^0,-\sigma_q^0)$ and define $\Omega = \mathrm{Diag}
(\sigma_p^2,- \sigma_q^2)$.  The involution $\tau_1(X) = - \Omega X^t
\Omega^{-1}$ together with $\eta_1=-1$ leads to
\begin{align*}
  &\mathfrak{k}\oplus \mathcal{Q}_+ = \Big\{
  \begin{pmatrix}
    A & 0 \\
    0 & D
  \end{pmatrix}
  \Big| A = \sigma_p^2 \bar{A} \sigma_p^2 \, ,\quad D=
  \sigma_q^2\bar{D} \sigma_q^2
  \Big\} ,\\
  & \mathfrak{p}= \Big\{
  \begin{pmatrix}
    0 & B \\
    B^\dagger & 0
  \end{pmatrix}
  \Big| B = - \sigma_p^2 \bar{B} \sigma_q^2 \Big\} , \quad
  \mathcal{Q}_-=\{s Y \mid Y\in \mathfrak{p}\} .
\end{align*}
A maximal Abelian subalgebra of $\mathfrak{h} \subset \mathcal{Q}_+$
is
\begin{align*}
  \mathfrak{h} = \{\mathrm{Diag}(\lambda_1, \ldots, \lambda_{p+q})
  \otimes \sigma^0 \mid \lambda_k \in \mathbb{R}\} .
\end{align*}
Note $s \in \mathfrak{h}$. Let $\lambda := \mathrm{Diag}(\lambda_1,
\ldots, \lambda_{p+q})\otimes \sigma^0$ and let $f_i \in
\mathfrak{h}^*$ be defined by $f_i(\lambda)= \lambda_i$. Then we have
\begin{align}
  &\Sigma_+(\mathfrak{k}\oplus \mathcal{Q}_+,\mathfrak{h})=\{f_k - f_l
  \mid 1\leq k < l\leq p \;\; \text{or} \;\; p < k < l \leq p+q \} ,\\
  &\Sigma_+(\mathfrak{p}\oplus \mathcal{Q}_-,\mathfrak{h}) = \{f_k -
  f_l \mid 1\leq k \leq p \;\; \text{and} \;\; p < l \leq p+q\} .
\end{align}
To determine the root multiplicities we note that the quaternions
$\{\sigma^0, i\sigma^1, i\sigma^2,i\sigma^3\}$ constitute a basis of
the space $\mathfrak{r} := \{X \in \mathfrak{gl}(2,\mathbb{C}) \mid X=
\sigma^2\bar{X} \sigma^2\}$. The root spaces corresponding to $f_k
-f_l$ then are
\begin{align*}
  &1 \leq k < l \leq p : \; \Big\{\begin{pmatrix} E_{kl} \otimes X &
    0 \\ 0 & 0 \end{pmatrix} \Big| X \in \mathfrak{r}\Big\} , \\
  &p < k < l \leq p+q : \; \Big\{\begin{pmatrix} 0 & 0 \\ 0 &
    E_{k-p,l-p} \otimes X \end{pmatrix} \Big| X \in \mathfrak{r}\Big\} ,
    \\&1\leq k \leq p < l\leq p+q: \; \Big\{\begin{pmatrix} 0 &E_{k,l-p}
    \otimes X \\ 0 & 0 \end{pmatrix} \Big| X \in \mathfrak{r}\Big\} .
\end{align*}
Thus all root spaces have dimension four and $J'$ is given by
\begin{align*}
    J'(\lambda)= \prod_{1\leq k < l \leq p+q } (\lambda_k - \lambda_l)^4 .
\end{align*}

\section{Proof} \label{Proof}

In the proof we use some standard results of Lie theory, all of which
can be found in the literature, e.g. in \cite{knapp05}.  Since
$\mathfrak{g}$ is closed under hermitian conjugation ($\dagger$) we
know that $\mathfrak{g}$ is reductive, i.e.\ the direct sum of an
Abelian and a semisimple Lie algebra. For simplicity we first
restrict ourselves to the case where $\mathfrak{g}$ is semisimple.
The extension to the reductive case will be straightforward.

The proof of the theorem is divided into three parts. The first part,
in section \ref{ps_domain}, contains the derivation of a new
parametrization of the PS domain, which makes it possible to deal
with its boundary. The second part, in \ref{ext_ps}, is concerned
with the extension of the PS domain to a domain without boundary.
First we identify good directions into which to extend the PS domain.
Then we give an extension of PS which does not change the value of
the integral. Although much of it is unnecessary for the formal
proof, section \ref{ext_ps} is an important prerequisite to
understanding the third part, \ref{equiv_PS_Euclid}, where we give a
homotopy $EPS$ connecting the extended PS domain to the Euclidean
domain. The main point is to make rigorous the following schematic
application of Stokes' theorem:
\begin{align*}
  \int_{PS} g(Q,A) & dQ = - \int_{EPS} \underbrace{d(g(Q,A) dQ)}_{=0}
  + \int_{Euclid} g(Q,A) dQ ,
\end{align*}
where we have introduced $g(Q,A):= e^{-\Tr(Q^2)-2i\Tr(Q A)}$. The
first term on the right hand side is identically zero because $g(Q,A)$
is holomorphic in $Q$. In the final subsection \ref{sect:4.4} we
deduce the corollaries \ref{cor_2_1} and \ref{cor_2_2}.

At this point a warning is in order. In the given form the expressions
above do not make sense. In order for the integrals over $PS$ and
$EPS$ to exist we have to include some regularization. This delicate
issue is discussed in detail in the last part of subsection
\ref{equiv_PS_Euclid}. That discussion also entails that the extension
of $PS$ does not contribute to the left hand side of the equation.

\subsection{A suitable parametrization of the PS domain}
\label{ps_domain}

We now invest some effort in order to derive a parametrization of the
domain of integration which gives full control over its boundary. To
guide the reader, we first define and explain all objects that are
necessary to formulate a lemma stating the parametrization.

In order to evaluate $e^Y X e^{-Y}$ in \eqref{ps_pk_param}
explicitly, we need to compute multiple commutators of $Y \in
\mathfrak{p}$ with $X \in \mathcal{Q}_+$. Therefore we now choose a
maximal Abelian subalgebra $\mathfrak{a}$ in $\mathfrak{p}$ and
diagonalize the commutator action of $\mathfrak{a}$ on $\mathcal{Q}$.
This diagonalization process gives rise to a root space decomposition
\begin{align*}
    \mathcal{Q} = \mathcal{Q}_0 \oplus \bigoplus_{\alpha \in
    \Sigma_+(\mathcal{Q}, \mathfrak{a})} (\mathcal{Q}_{\alpha} \oplus
    \mathcal{Q}_{-\alpha}) ,
\end{align*}
where $\Sigma_+(\mathcal{Q}, \mathfrak{a})$ denotes a set of positive
roots. Each root space in turn is decomposed into a hermitian
($\mathcal{Q}_+$) and an antihermitian ($\mathcal{Q}_-$) part:
\begin{align*}
  \mathcal{Q}_{\pm , \alpha} := \Fix_{\pm \theta}
  (\mathcal{Q}_{\alpha} \oplus \mathcal{Q}_{-\alpha}) \subset
  \mathcal{Q}_{\pm} \, .
\end{align*}
We also let $\mathcal{Q}_{\pm,0 } := \Fix_{\pm \theta}
(\mathcal{Q}_{0} ) \subset \mathcal{Q}_{\pm}$. Hence we have the
decompositions
\begin{align} \label{Q_decomposition}
  \mathcal{Q}_{\pm}= \mathcal{Q}_{\pm,0} \oplus \bigoplus_{\alpha \in
    \Sigma_+(\mathcal{Q},\mathfrak{a})} \mathcal{Q}_{\pm, \alpha}\, .
\end{align}
For future reference we observe that
\begin{align}
  \ad(s) :\; \mathfrak{p} \rightarrow \mathcal{Q}_- \,,
  \quad Y \mapsto [s,Y] , \label{ad_s_isom}
\end{align}
is an isomorphism. This fact will be used several times in the proof.

For the following constructions we review the notion of pointed
polyhedral cone and triangulations thereof \cite{zieg94, deloera10}.
A pointed polyhedral cone is a subset of a vector space. By
definition it is an intersection of finitely many half spaces where
the intersection of all hyperplanes bounding the half spaces contains
only the zero vector. The word pointed reflects the fact that there
exists a hyperplane which intersects the cone only at zero, with the
rest of the cone lying strictly on one side of that hyperplane. For
example, if $\Sigma_+(\mathfrak{g},\mathfrak{a})$ denotes a system of
positive roots for the adjoint action of $\mathfrak{a}$ on
$\mathfrak{g}$, the positive Weyl chamber
\begin{align*}
  \mathfrak{a}^+ = \bigcap_{\beta \in \Sigma_+(\mathfrak{g},
    \mathfrak{a})} \{H \in \mathfrak{a} \mid \beta(H) \geq 0 \}
\end{align*}
is a pointed polyhedral cone. In the following we refer to a pointed
polyhedral cone as a cone for short.

Let $E \subset \mathfrak{a}$ be a vector space of codimension one
such that $\mathfrak{a}^+$ lies entirely on one side of $E$. A face
of $\mathfrak{a}^+$ is a set of the form $\mathfrak{a}^+ \cap E$. The
zero vector is the unique zero dimensional face. It is convenient
also to include the empty set as a face. The one dimensional faces
are called edges. Note that each nontrivial face is again a cone.

It is a fact \cite{zieg94} that any cone $\mathfrak{a}^+$ admits a
different representation: there exist $m$ elements $H_i' \in
\mathfrak{a}$ such that
\begin{align*}
  \mathfrak{a}^+ = \left\{\sum_{i=1}^m h^i H_i' \mid h^i \geq 0
  \right\} .
\end{align*}
The $H_i'$ are called generators of the cone. They can be chosen in
such a way that each $H_i'$ generates an edge of the cone. In the
case of a positive Weyl chamber $\mathfrak{a}^+$ a set of generators
is furnished by the simple co-roots. A cone is called simplicial if
its generators are linearly independent. A $d$-cone is a cone of
dimension $d$. It is a known fact of Lie theory that $\mathfrak{a}^+$
is a simplicial $\text{dim} (\mathfrak{a})$-cone.

A finite collection $T$ of $\text{dim}(\mathfrak{a})$-cones is called
a subdivision of $\mathfrak{a}^+$ if $\mathfrak{a}^+ = \cup_{S \in T}
S$ and $S_1 \cap S_2$ is a face of both $S_1$ and $S_2$ for all $S_1,
S_2\in T$. If each cone in a subdivision $T$ is simplicial, then $T$
is called a triangulation.

Bearing these facts in mind we proceed to describe a decomposition of
$\mathfrak{a}^+$ which, as we shall see below, is directly related to
the boundary of the PS domain. Note, first of all, that a root
$\alpha \in \Sigma_+(\mathcal{Q},\mathfrak{a})$ may change sign on
$\mathfrak{a}^+$ since $\mathfrak{a}^+$ is defined with respect to
the root system $\Sigma_+(\mathfrak{g},\mathfrak{a})$. The closures
of the connected components of $\mathfrak{a}^+ \setminus \left(
\mathfrak{a}^+ \cap (\cup_{\alpha } \ker(\alpha)) \right)$ can be
obtained as appropriate intersections of half spaces and hence are
again cones. Let $\{H_i \}_{i = 1, \ldots, M }$ denote the collection
of generators of these cones [the cardinality $M$ exceeds $m$ if
$\mathfrak{a}^+ \cap (\cup_{\alpha } \ker(\alpha)) \neq \emptyset$].
By construction the intersection of two such cones is a face common
to both. Put differently, the generators common to two such cones
generate a joint face. Thus the decomposition we have just described
yields a subdivision of $\mathfrak{a}^+$. It is a fact
\cite{zieg94,deloera10} that every subdivision of a cone can be
refined to a triangulation without introducing any new generators.

For the rest of the article we fix a triangulation
\begin{align} \label{cone_dec}
  \mathfrak{a}^+ = \bigcup_{c \in C} \mathfrak{a}^{+}_{c} \,
\end{align}
which refines the subdivision of $\mathfrak{a}^+$ described above.
Let $I_c\subset \{1,\ldots, M\}$ be such that $\{H_i\}_{i \in I_c}$
is the set of generators for the simplicial cone indexed by $c\in C$,
i.e., let
\begin{align*}
  \mathfrak{a}_c^+= \left\{\sum\nolimits_{i\in I_c} h^i H_i
  \mid h^i \geq 0 \right\} .
\end{align*}
Note that $| I_c | = \mathrm{dim}\, \mathfrak{a}$ and that the
generators $\{H_i\}_{i \in I_c}$ form a basis of $\mathfrak{a}$. The
latter fact implies that each $H \in \mathfrak{a}^{+}_{c}$ is
represented uniquely as
\begin{align} \label{cone_rep}
  H =\sum_{i \in I_c} h^i H_i
\end{align}
with coefficients $h^i \in \mathbb{R}^+$.  The intersection
$\mathfrak{a}_c^+ \cap \mathfrak{a}_{c'}^+$ of two simplicial cones
is again a simplicial cone; indeed, the set of generators of the
latter is $\{H_i\}_{i \in I_c \cap I_{c'}}$. A key property of the
decomposition \eqref{cone_dec} is that the sign of each $\alpha\in
\Sigma_+(\mathcal{Q},\mathfrak{a})$ stays constant on any given
simplicial cone $\mathfrak{a}_c^+$. However it may still happen that
$\alpha$ vanishes on the boundary of $\mathfrak{a}_c^+$.

\begin{figure}
\centering
\includegraphics{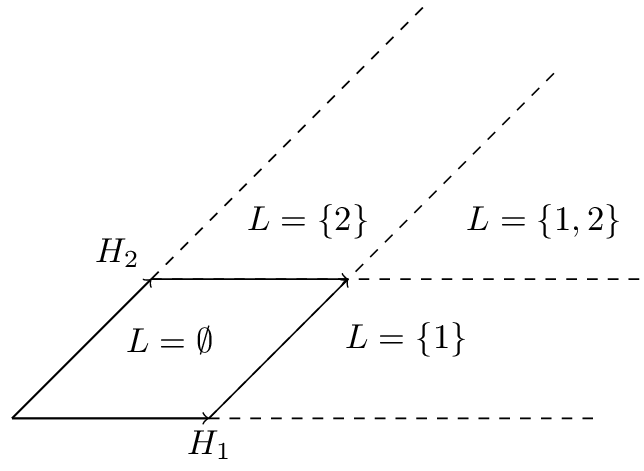}
\caption{This figure shows $\mathfrak{a}^+$ for
  $\mathfrak{g}=\mathfrak{su}(2,2)$. The index $c$ has been omitted
  since there exists only one simplicial cone in this case. The
  possible $L$'s are subsets of $\{1,2\}$.}
\label{eps_domains}
\end{figure}

Next we introduce a subdecomposition of each cone $\mathfrak{a}_c^+$.
Let $L\subset I_c$ and define
\begin{align*}
  \mathfrak{a}^+_{L,c}:= \Big\{\sum_{i\in I_c} h^i H_{i} \in
  \mathfrak{a}^+_c \mid \forall i \in L :\, h^i \geq 1 \;\;
  \mathrm{and}\;\; \forall i \notin L :\, h^i \leq 1 \Big\} .
\end{align*}
An example of this decomposition is shown in figure
\ref{eps_domains}. It may be a helpful observation to note that the
decomposition
\begin{align*}
    \mathfrak{a}^+ = \bigcup_{c \in C} \bigcup_{L \subset I_c}
    \mathfrak{a}^{+}_{L,c}
\end{align*}
carries the structure of a simplicial complex.

With these definitions understood we introduce for each index pair
$(\alpha, c)$ a function on $\mathfrak{a}$ by
\begin{align*}
    T_{\alpha,c} :\; \mathfrak{a} &\rightarrow \mathbb{R} , \\
    H = \sum_{i\in I_c} h^i H_{i} &\mapsto \left\lbrace
    \begin{array}{ll} \tanh\Big(\sum_{i\in I_c} \frac{h^i}{1-h^i}
    \alpha(H_{i})\Big), & \forall i \in I_c : \, \big( h^i < 1 \;
    \text{or} \;\alpha(H_{i}) = 0 \big) , \\
    \sgn(\alpha(H_{i})), & \text{else} ,\end{array} \right.
\end{align*}
where $H = \sum_{i \in I_c} h^i H_i$ is meant in the sense of
\eqref{cone_rep} with coefficients $h^i \in \mathbb{R}$. In order for
this function $T_{\alpha,c}$ to be well-defined it is crucial that
the decomposition of $\mathfrak{a}^+$ into simplicial cones is such
that for fixed $c$ and fixed $\alpha$ the sign of $\alpha(H_{i})$ is
the same for all $i\in I_c$ with $\alpha(H_{i})\neq 0$.

We are now going to formulate a lemma which summarizes what we are
aiming at in this section. For that purpose we introduce $K :=
\exp(\mathfrak{k})$ and let $Z_K(\mathfrak{a})$ be the centralizer of
$A = \exp(\mathfrak{a})$ in $K$. Fixing some $H \in \mathfrak{a}$ with
$\alpha(H) \neq 0$ for all $\alpha \in \Sigma_+(\mathcal{Q},
\mathfrak{a})$ we define
\begin{align*}
 \phi : \; \mathcal{Q}_{\alpha} \oplus \mathcal{Q}_{-\alpha} &\rightarrow
 \mathcal{Q}_{\alpha} \oplus \mathcal{Q}_{-\alpha} \, ,\\
 Z + Z' &\mapsto \alpha(H)^{-1} [H, Z+Z'] = Z-Z' .
\end{align*}
Note that $\phi$ satisfies
\begin{align*}
 \phi \circ \phi = id \; \; \mathrm{and} \; \;\phi(\mathcal{Q}_{\pm,
   \alpha}) = \mathcal{Q}_{\mp,\alpha} \, .
\end{align*}
In addition we define orthogonal projections
\begin{align*}
\pi_{\pm,\alpha}: \mathcal{Q} \rightarrow \mathcal{Q}_{\pm, \alpha} \, .
\end{align*}

The following lemma contains a parametrization of the domain of
integration which gives direct control over its boundary.
\begin{lem} \label{lem_parametrisation}
The mappings
\begin{align} \label{PS_c_parametrisation}
  PS_c : \; &\mathfrak{a}^+_{\emptyset, c} \times \left( K
    \times_{Z_K(\mathfrak{a})}\mathcal{Q}_+\right) \rightarrow
  \mathcal{Q} ,\\ &(H, [k;X]) \mapsto \Ad(k) \Big(X_0 +
  \sum_{\alpha \in \Sigma_+(\mathcal{Q},\mathfrak{a})} (X_{\alpha} +
  T_{\alpha,c}(H) \, \phi( X_{\alpha})) \Big) ,\notag
\end{align}
with $X_{\alpha} := \pi_{+, \alpha} (X)$ and $[k;X]=
[kz^{-1};zXz^{-1}]$ for $z \in Z_{K}(\mathfrak{a})$ have the following
properties:
\begin{enumerate}[label=\roman{*})]
\item \label{lem_par_it_boundary} The boundary $\partial PS_c$ (in the
  sense of integration chains) is obtained by applying the boundary
  operator $\partial$ to $\mathfrak{a}^+_{\emptyset, c}$.
\item \label{lem_par_it_orientation} A choice of orientation on
  $\mathfrak{a}^+ \times \left(K \times_{Z_K(\mathfrak{a})}
    \mathcal{Q}_+ \right)$ induces an orientation for each
  $PS_c$. There exists a particular choice of orientation for which
  $PS = \sum_{c \in C} PS_c$ holds, where the equality sign is meant
  in the sense of integration chains.
\item \label{lem_par_it_codim}The contributions to the boundary
    of $PS_c$ which come from $\partial \mathfrak{a}^+_{
    \emptyset, c} \cap \partial \mathfrak{a}^+$ are of
    codimension at least two and can be neglected.
\end{enumerate}
\end{lem}

To prove lemma \ref{lem_parametrisation} we perform a sequence of
four reparametrizations of the original parametrization $PS$. The
first three reparametrizations are preparatory and do not relate
directly to lemma \ref{lem_parametrisation}. Each reparametrization
is discussed in a separate subsection for clarity.

\subsubsection{Reparametrization I: Decomposition of $\mathfrak{p}$}

The goal of the next three reparametrizations is to evaluate
$PS(Y,X)=\Ad(e^Y)X$ in more detail. Key to this is a choice of maximal
Abelian subalgebra $\mathfrak{a} \subset \mathfrak{p}$ whose adjoint
action on $\mathcal{Q}=\mathcal{Q}_+ \oplus \mathcal{Q}_-$ is
diagonalizable. To get started, we parametrize $\mathfrak{p}$ using
$K/Z_K(\mathfrak{a})$ and the interior $(\mathfrak{a}^+)^o$ of
$\mathfrak{a}^+$:
\begin{align*}
  R_{\rm I} : \; (\mathfrak{a}^+)^o \times K/Z_{K}(\mathfrak{a})
  &\rightarrow \mathfrak{p} ,\\ (H, [k]) & \mapsto k H k^{-1} .
\end{align*}
$R_{\rm I}$ is obviously well defined, and it is a standard fact that
$R_{\rm I}$ is injective for semisimple Lie algebras with Cartan
decomposition $\mathfrak{g}=\mathfrak{k} \oplus \mathfrak{p}$. Hence
$R_{\rm I}$ is a diffeomorphism onto $\mathrm{Im}(R_{\rm I})$. Note that
$\mathfrak{p} \setminus \mathrm{Im}(R_{\rm I}) $ is a set of measure zero
since $\mathfrak{p}= \cup_{k \in K} k \mathfrak{a}^+ k^{-1}$ (see
e.g. \cite{knapp05}) and
\begin{align*}
  \mathrm{Im}(R_{\rm I}) = \cup_{k \in K} k \left(\mathfrak{a}^+ \setminus
  (\mathfrak{a}^+ \cap (\cup_{\alpha} \ker \alpha))\right) k^{-1},
\end{align*}
where $\alpha$ runs over the roots in $\Sigma_+(\mathfrak{g},
\mathfrak{a})$.

Precisely speaking, we are going to use the parametrization
\begin{align*}
  PS \circ R_{\rm I} : \; \mathfrak{a}^+ \times K/Z_{K}(\mathfrak{a})
  \times \mathcal{Q}_+ \rightarrow \; & \mathcal{Q} , \\
  (H,[k], X) \mapsto \; & e^{kHk^{-1}} X e^{-k Hk^{-1}} .
\end{align*}
Recall that the PS domain is oriented by an orientation of
$\mathfrak{p}\oplus \mathcal{Q}_+$. Declaring $R_{\rm I}$ to be
orientation preserving induces an orientation on $\mathfrak{a}^+
\times K/Z_{K}(\mathfrak{a}) \times \mathcal{Q}_+$.

Further reparametrizations of the PS domain are introduced below. To
avoid an overload of notation, we will denote each new
parametrization still by $PS$.

\subsubsection{Reparametrization II: Twisting $K/{Z_K(\mathfrak{a})}$
and $\mathcal{Q}_+$}

In this section we prepare the further evaluation of the
$\ad(\mathfrak{a})$ action in the next subsection. Consider the
reparametrization
\begin{align*}
  R_{\rm II} : \; K \times_{Z_K(\mathfrak{a})} \mathcal{Q}_+
  &\rightarrow K/Z_K(\mathfrak{a}) \times \mathcal{Q}_+ \, ,
  \\ [k z^{-1};z X z^{-1}]&\mapsto ([k], kXk^{-1}) ,
\end{align*}
where the expression $[k; X]\equiv [k z^{-1}; z X z^{-1}]$ (for $z \in
Z_{K}(\mathfrak{a})$) stands for an equivalence class of the group
action of $Z_K(\mathfrak{a})$ on $K \times \mathcal{Q}_+$. This group
action defines the trivial bundle $K \times_{Z_K(\mathfrak{a})}
\mathcal{Q}_+$. The inverse of $R_{\rm II}$ is
\begin{align*}
  R_{\rm II}^{-1} : \; K/Z_K(\mathfrak{a}) \times \mathcal{Q}_+
  &\rightarrow K \times_{Z_K(\mathfrak{a})} \mathcal{Q}_+ \, ,\\
  ([k],X)& \mapsto [k; k^{-1}Xk] .
\end{align*}
$R_{\rm II}$ is a diffeomorphism and can therefore be used as a
reparametrization to obtain the new parametrization
\begin{align}
  PS \circ R_{\rm II} : \; \mathfrak{a}^+ \times K \times_{Z_{K}
  (\mathfrak{a})} \mathcal{Q}_+ \rightarrow \; & \mathcal{Q} , \notag\\
  (H, [k z; z X z^{-1}]) \mapsto \; & e^{k H k^{-1}} k X k^{-1} e^{-kH
  k^{-1}} \notag \\ &= k e^{H} X e^{-H} k^{-1} = \Ad(k) (e^{\ad(H)}
  X) . \label{PSII}
\end{align}

\subsubsection{Decomposition of $\mathcal{Q}_+$}

Since $Z_{K}(\mathfrak{a})$ is a subgroup of $K$ and, by definition,
commutes with the $\ad(\mathfrak{a})$ action on $\mathcal{Q}$, the
parametrization \eqref{PSII} is compatible with the decomposition
\eqref{Q_decomposition} of $\mathcal{Q}$. In particular we have
\begin{align*}
  \pi_{+,\alpha}\circ \Ad(z) = \Ad(z)\circ \pi_{+,\alpha}
\end{align*}
for $z \in Z_{K}(\mathfrak{a})$ and $\alpha \in \Sigma_+(\mathcal{Q},
\mathfrak{a})$. A short calculation for $X_{\alpha}\in \mathcal{Q}_{
+,\alpha}$ gives
\begin{align*}
  e^{\ad(H)} X_{\alpha} &= \cosh(\alpha(H)) X_{\alpha} +
  \sinh(\alpha(H)) \phi(X_{\alpha}) .
\end{align*}
Hence the parametrization \eqref{PSII} can be rewritten as
\begin{align}
  PS: \; &\mathfrak{a}^+ \times K \times_{Z_{K}(\mathfrak{a})}
  \mathcal{Q}_+ \rightarrow  \mathcal{Q} , \notag\\ (H,[k;&X])
  \mapsto \Ad(k) \Big( X_0 + \sum_{\mathclap{\alpha \in
  \Sigma_+(\mathcal{Q},\mathfrak{a})}} \big(\cosh(\alpha(H))
  X_{\alpha} + \sinh(\alpha(H)) \phi(X_{\alpha})\big) \Big) ,
  \label{PS_decomp}
\end{align}
where $X_{\alpha} = \pi_{+,\alpha}(X)$.

\subsubsection{Reparametrization III: Rectification}

\begin{figure}
\centering
\includegraphics{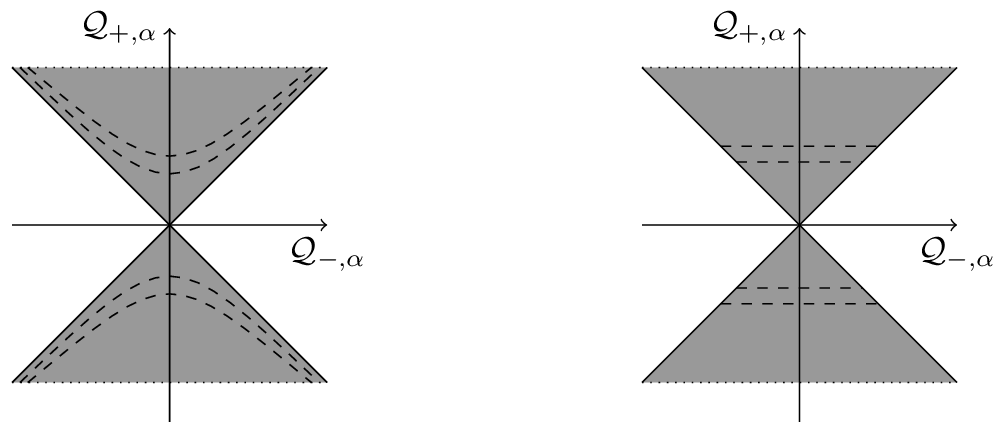}
\caption{Motivation for the third reparametrization step. The dashed
  lines are the images of straight lines through the origin in
  $\mathfrak{a}^+$ before (left) and after (right) the
  reparametrization $R_{\rm III}$. }
\label{rep_3}
\end{figure}

As a motivation for the next reparametrization we note that $(X,Y)
\mapsto \Tr(X Y)$ is an $\Ad(K)$-invariant scalar product on
$\mathcal{Q}$ and that all the different spaces
$\mathcal{Q}_{\pm,\alpha}$ are orthogonal to each other. For the
moment, we fix $\alpha \in \Sigma_+(\mathcal{Q}, \mathfrak{a})$ and
consider only the part
\begin{align*}
  \cosh(\alpha(H)) X_{\alpha} + \sinh(\alpha(H)) \phi(X_{\alpha})
\end{align*}
of the parametrization \eqref{PS_decomp}. The corresponding
two-dimensional picture is shown in figure \ref{rep_3}, where we see
the image of a straight line through the origin in $\mathfrak{a}^+$ as
a hyperbola.
We are going to change the parametrization in such a way that the
hyperbola is rectified to a straight line; see figure \ref{rep_3}.
Such a reparametrization gives us a handle on the boundary of the PS
domain, as is discussed in the next subsection. Accordingly, the
third reparametrization we use is given by
\begin{align*}
  R_{\rm III} : \; \mathfrak{a}^+ \times K \times_{Z_{K}(\mathfrak{a})}
  \mathcal{Q}_+ &\rightarrow \mathfrak{a}^+ \times K
  \times_{Z_{K}(\mathfrak{a})} \mathcal{Q}_+ \, , \\
  (H,[k;X]) &\mapsto \Big(H,\Big[k;X_0 +\sum_{\alpha \in
    \Sigma_+(\mathcal{Q},\mathfrak{a})}\frac{1}{\cosh(\alpha(H))}
  X_{\alpha}\Big]\Big) .
\end{align*}
This is another orientation preserving diffeomorphism. We thus obtain
\begin{align}
  PS \circ R_{\rm III} : \; &\mathfrak{a}^+ \times K
  \times_{Z_{K}(\mathfrak{a})}
  \mathcal{Q}_+ \rightarrow  \mathcal{Q} ,\notag\\
  (H,&[k;X]) \mapsto \Ad(k) \Big( X_0 + \sum_{\mathclap{\alpha \in
      \Sigma_+(\mathcal{Q},\mathfrak{a})}} \big( X_{\alpha} +
  \tanh(\alpha(H)) \phi(X_{\alpha})\big) \Big) , \label{ps_param}
\end{align}
which is renamed to $PS$ in the following.

\subsubsection{Reparametrization IV: Making the boundary visible}

{}From the parametrization \eqref{ps_param} (see also figure
\ref{rep_3}) it is clear that the boundary is reached when some
$\alpha(H)$ goes to $\pm \infty$ and hence $\tanh$ goes to $\pm 1$.
Put differently, the boundary of the domain of integration can be
reached through a limit in the parameter space $\mathfrak{a}^+$. To
obtain control over the boundary we have to make sense of the
expression $\lim_{H\rightarrow \infty}\tanh(\alpha(H))$. This limit
is encoded in the functions $T_{\alpha, c}$. Recall that
\begin{align*}
  T_{\alpha,c}\left( \sum_{i\in I_c} h^i H_{i} \right) =
  \left\lbrace \begin{array}{l l}\tanh\Big(\sum_{i\in I_c}
      \frac{h^i}{1-h^i} \alpha(H_{i})\Big),
      & \forall i: \, h^i < 1\; \text{or} \;\alpha(H_{i}) = 0 ,\\
      \sgn(\alpha(H_{i})), & \text{else} ,\end{array} \right.
\end{align*}
where the index $c$ refers to the decomposition \eqref{cone_dec} of
$\mathfrak{a}^+$ into the simplicial cones $\mathfrak{a}^+_c$. Fix
$\alpha \in \Sigma_+(\mathcal{Q},\mathfrak{a})$ and $j\in I_c$ such
that $\alpha(H_{j})\neq 0$. Then
\begin{align*}
  \lim_{h^j \rightarrow 1} T_{\alpha, c}\left(\sum_{i\in I_c} h^i
    H_{i}\right)= \sgn(\alpha(H_{j})) .
\end{align*}
This shows that $T_{\alpha,c}$ is continuous. $T_{\alpha,c}$ is also
differentiable since
\begin{align*}
  \lim_{h^j \rightarrow 1} \partial_{h^j}\tanh\left(\sum_{i\in I_c}
    \frac{h^i}{1-h^i}\alpha(H_{i})\right) =0
\end{align*}
generalizes to all higher (and mixed) partial derivatives.

\begin{figure}
\centering
\includegraphics{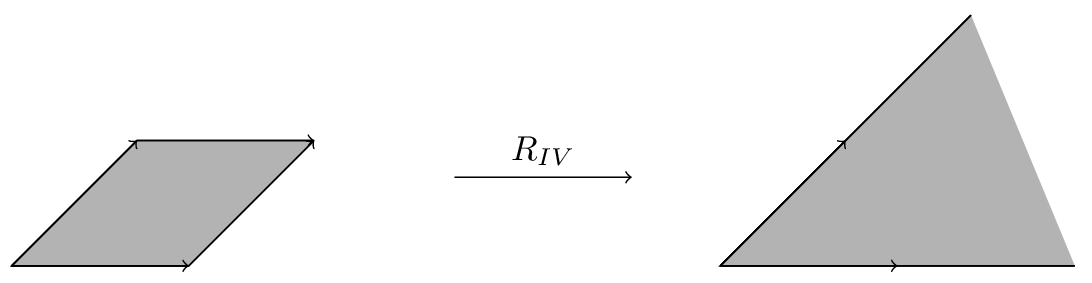}
\caption{$\mathfrak{su}(2,2)$ example for $R_{\rm IV}$ mapping
  $\mathfrak{a}_{\emptyset}^+$ on the left hand side to
  $\mathfrak{a_+}$ on the right side. In this example there is only
  one simplicial cone.}
\label{rep_4}
\end{figure}

To put the functions $T_{\alpha, c}$ to use we define for each cone
$c\in C$ the mapping
\begin{align*}
  R_{{\rm IV},c}: \; (\mathfrak{a}_{\emptyset,c}^+)^o\times K
  \times_{Z_{K}(\mathfrak{a})} \mathcal{Q}_+ &\rightarrow
  \mathfrak{a}_{c}^+\times K \times_{Z_{K}(\mathfrak{a})}
  \mathcal{Q}_+ \, ,\\
  \left(\sum_{i\in I_c} h^i H_{i},[k;X] \right)&\mapsto
  \left(\sum_{i\in I_c} \frac{h^i}{1-h^i} H_{i},[k;X] \right) ,
\end{align*}
where $(\mathfrak{a}_{\emptyset,c}^+)^o$ denotes the interior of
$\mathfrak{a}_{\emptyset,c}^+$. For each simplicial cone, this is an
(orientation preserving) diffeomorphism onto its image. The mapping
is visualized in figure \ref{rep_4}. We use it to reparametrize $PS$
on each cone. We thus obtain $PS_c = PS\circ R_{{\rm IV},c}$ , which
is the parametrization defined in lemma \ref{lem_parametrisation},
eq.\ $\eqref{PS_c_parametrisation}$.

In the following we want to give the notion `boundary of the PS
domain' a precise meaning. In the case of integration cells, i.e.,
differentiable mappings defined on a cube, the boundary operator
$\partial$ is defined as usual. $\partial$ can also be applied to
integration chains, i.e.\ formal linear combinations of cells. In
principle the correct procedure would be to decompose each $PS_c$
into cells in order to apply $\partial$. However, in the following we
argue that we can treat each $PS_c$ effectively as single cell with
the boundary operator $\partial$ acting just on the $\mathfrak{a}_{
\emptyset,c}^+$ part of the domain of definition. Note that
$\mathfrak{a}_{\emptyset,c}^+$ by the decomposition \eqref{cone_rep}
is diffeomorphic to an $n$-dimensional cube.

First note that $PS_c$ extends as a differentiable mapping to a
neighborhood of $\mathfrak{a}_{\emptyset,c}^+$ since the
$T_{\alpha,c}$ are differentiable functions defined on
$\mathfrak{a}$. Thus it is possible to define the orientation of the
boundary. Furthermore, since $K$ is a closed compact manifold it
suffices to discuss boundary contributions arising from a
decomposition of $\mathcal{Q}_+$ into cells. By inspecting our
parametrization we see that going to infinity on $\mathcal{Q}_+$
implies going to infinity in the domain of integration. In section
\ref{equiv_PS_Euclid} we show that the integrand goes to zero
exponentially on this domain and hence all possible boundary
contributions vanish. Thus we obtain part \ref{lem_par_it_boundary}
of lemma \ref{lem_parametrisation}

Part \ref{lem_par_it_orientation} of lemma \ref{lem_parametrisation}
follows immediately by noting that the mappings $R_{{\rm IV},c}$ are
diffeomorphisms onto.

Since similar arguments are needed in several other parts of section
\ref{Proof}, the proof of part \ref{lem_par_it_codim} of lemma
\ref{lem_parametrisation} is presented in appendix
\ref{app_contributions}.

\subsection{Extending the PS domain}\label{ext_ps}

In this section we construct an extension of the PS domain which has
no boundary other than the irrelevant boundary at infinity. The idea
is to connect each boundary point to infinity by attaching one
halfline. The directions of these halflines should be such that the
attached domain does not contribute to the integral of $g(Q,A)
\chi_{\epsilon}(Q)\,dQ$.  We first determine such a direction for
each boundary point, and then give a parametrization of the attached
domains. In the following let
\begin{align}
  B(X,Y):=\Tr(XY)
\end{align}
denote the trace form on $\mathcal{Q}^{\mathbb{C}}$.

Some care must be exercised in order to guarantee the convergence of
the integral on the extended PS domain. The positivity requirement
$As>0$ in the theorem already gives a hint that the matrix $s$ plays a
prominent role in the discussion of convergence. Owing to $s \in
\mathcal{Q}_+$ we have the decomposition
\begin{align} \label{s_decomposition}
  s= s_0 + \sum_{\mathclap{\alpha \in
      \Sigma_+(\mathcal{Q},\mathfrak{a})}} s_{\alpha} \, ,
\end{align}
where $s_0 \in \mathcal{Q}_{+,0}$ and $s_{\alpha} = \pi_{+,\alpha}(s)
\in \mathcal{Q}_{+,\alpha}$.

The next lemma introduces the convergent directions which are used to
extend $PS$ in such a way that convergence is maintained.

\begin{lem} \label{E_lem}
The matrices
\begin{align} \label{def_E}
  E_{j} := 2 \lim_{t \rightarrow \infty} \frac{\Ad(e^{t
      H_{j}})s}{\mathrm{max}_{\alpha \in
      \Sigma_+(\mathcal{Q},\mathfrak{a})} \; e^{|\alpha(t H_{j})|}}
\end{align}
are well defined and non-zero. The following properties hold for all
$1\leq i,j\leq M$:
\begin{enumerate}[label=\roman{*})]
\item \label{inequality}$\Tr(E_{i} \Ad(k)^{-1} A) > 0$.
\item There exist numbers $e_{j}^{\alpha}\in \{0,1\}$ such that
    the matrices $E_{j}$ decompose as
\begin{align}\label{E_{i,c}_explicit_decomposition}
  E_{j}= \sum_{\alpha(H_{j})\neq 0} e_{j}^{\alpha} \big(
  s_{\alpha} + \sgn(\alpha(H_{j})) \phi(s_{\alpha}) \big) .
\end{align}
\item For $i,j \in I_c$ one has $B(E_{i}, E_{j}) = 0$. In
    particular, $B(E_{i}, E_{i}) = 0$.
\end{enumerate}
\end{lem}
Before we come to the proof of lemma \ref{E_lem}, we formulate another
lemma which suggests how to extend the PS domain.  Note however that
an integral over $PS$ needs a regulating function and that we postpone
the discussion of convergence to section \ref{equiv_PS_Euclid}.
Strictly speaking, the next lemma is not necessary for the proof of
theorem \ref{thm_2_1}. It is included as a preparation for the more
involved definition of the homotopy introduced in section
\ref{equiv_PS_Euclid}.

\begin{lem}\label{Extension_lem}
For $c\in C$ and $L \subset I_c$ the mappings
\begin{align*}
  PS_{L,c} : \; &\mathfrak{a}_{L,c}^+ \times K
  \times_{Z_K(\mathfrak{a})} \mathcal{Q}_+ \rightarrow
  \mathcal{Q}^{\mathbb{C}} ,\\ (H,[k; &X]) \mapsto \Ad(k) \Big( X_0 +
  \sum_{\mathclap{\alpha \in \Sigma_{+}(\mathcal{Q},\mathfrak{a}) }}
  \Big(X_{\alpha} + T_{\alpha,c}(H) \phi(X_{\alpha})\Big) -i \sum_{j
    \in L} (h^j-1) E_{j} \Big) ,
\end{align*}
are well defined as integration chains and one has
\begin{align} \label{PS_boundary}
  \partial \left( \sum_{c\in C} \sum_{ L \subset I_c} PS_{L,c}
  \right)= 0 ,
\end{align}
as long as the sum of chains is integrated against forms with
sufficiently rapid decay at infinity.
\end{lem}

\subsubsection{Proof of lemma \ref{E_lem}}

To see that the matrices $E_{i}$ are well defined, we express them in
a more explicit fashion. A short calculation using
\eqref{s_decomposition} gives
\begin{align*}
  \Ad(e^{H})s = s_0 + \sum_{\mathclap{\alpha \in
      \Sigma_+(\mathcal{Q},\mathfrak{a})}} \big(\cosh(\alpha(H))
  s_{\alpha} + \sinh(\alpha(H))\phi(s_{\alpha}) \big) .
\end{align*}
This shows that the limit in \eqref{def_E} exists and that the
matrices $E_j$ decompose as shown in equation
\eqref{E_{i,c}_explicit_decomposition}. Thus we obtain
$ii)$. Recalling that $\ad(s)|_{\mathfrak{p}}$ is injective we
conclude that each of the matrices $E_j$ is non-zero.

For property $i)$ we note that $As > 0$ and
\begin{align*}
  s^{-1} E_{j} = 2 \lim_{t\rightarrow \infty}\frac{e^{-2tH_{j}}}
  {\mathrm{max}_{\alpha \in \Sigma_+(\mathcal{Q},\mathfrak{a})}
  \; e^{|\alpha(t H_{j})|}} \geq 0 .
\end{align*}
Thus $\Tr((\Ad(k)s^{-1} E_{j}) A s )$ is the trace of the product of
the non-zero positive semidefinite hermitian matrix $s^{-1}\Ad(k)E_j$
and the positive hermitian matrix $As$. By inserting the eigenvalue
representation $\sum_n p_n \pi_n$ of $s^{-1}\Ad(k)E_j$ we get
\begin{align*}
  \Tr(s^{-1} \Ad(k)E_{j} A s )= \sum_n p_n \Tr(\pi_n As) .
\end{align*}
Since $As$ is positive we have $\Tr(\pi_n As) > 0$. Property $i)$ then
follows because $s^{-1}\Ad(k)E_j\neq 0$ implies that there exists some
$p_n > 0$.

To prove property $iii)$ we use $ii)$ and note that for all $\alpha,
\beta \in \Sigma_+(\mathcal{Q},\mathfrak{a})$ we have the
orthogonality relations
\begin{align*}
  B(s_{\alpha}, s_{\beta}) = - B(\phi(s_{\alpha}), \phi(s_{\beta}))
  = \delta_{\alpha, \beta} B(s_\alpha , s_\alpha)
\end{align*}
and $B(s_{\alpha}, \phi(s_{\beta})) = 0$. We also recall that a fixed
root $\alpha$ does not change sign on a fixed simplicial cone
$\mathfrak{a}_c^+$. The desired result $B(E_i , E_j) = 0$ for $i,j \in
I_c$ then follows directly.

\subsubsection{Proof of lemma \ref{Extension_lem}}

In this section we show that the mappings $PS_{L,c}$ in lemma
\ref{Extension_lem} have the stated properties. First of all, the
mappings are well defined since $Z_{K}(\mathfrak{a})$ acts trivially
on the matrices $E_{i}$.  We recall that the situation is visualized
in figure \ref{eps_domains}.

$PS_{L,c}$ can be extended as a mapping to $\mathfrak{a}$ since
$T_{\alpha,c}$ makes sense on $\mathfrak{a}$ and so does $(h^j-1)$.
Thus $PS_{L,c}$ is well defined as an integration chain. By the same
argument as for the $PS_{\emptyset,c}$ case, a boundary can arise
only by the action of the boundary operator $\partial$ on the factor
$\mathfrak{a}_{L,c}^+$.

In the following we always neglect possible boundary contributions
from infinity, since the integrals under consideration are convergent
by assumption.

To see that the different integration cells $PS_{L,c}$ and $PS_{L',
c'}$ fit together in a seamless way, we recall that $\mathfrak{a}_c^+
\cap \mathfrak{a}_{c'}^+$ is again a simplicial cone which is
generated by the set $\{H_i\}_{i \in I_c \cap I_{c'}}$. In particular
each $H \in \mathfrak{a}_{L,c}^+ \cap \mathfrak{a}_{L',c'}^+$ can be
represented in the form $H=\sum_{i \in I_c \cap I_{c'}} h^i H_i$,
which implies that $T_{\alpha,c}(H) = T_{\alpha, c'}(H)$. We have
$h^i\geq 1$ for $i\in L$ and $h^i \leq 1$ for $i \in I_c \setminus
L$. Together with similar conditions from $L'$ this yields
\begin{align*}
  \sum_{i \in L} (h^i -1) E_i = \sum_{i \in L'} (h^i -1) E_i
\end{align*}
on the intersection $\mathfrak{a}_{L,c}^+ \cap \mathfrak{a}_{L',
c'}^+$. Hence we obtain the equality $PS_{L,c}(H, [k;X]) =
PS_{L',c'}(H, [k;X])$ on the joint domain of definition.

Moreover, the induced orientations on the boundaries between two
neighboring cells are opposites of each other. Together with the fact
(shown in appendix \ref{app_contributions}) that the contributions
from $\partial \mathfrak{a}^+$ are of codimension no less than two,
these results yield \eqref{PS_boundary}.

To get some intuition for the situation it is useful to observe that
the halflines which are glued to boundary points of the PS domain,
point into directions within $\oplus_{\alpha \in \Sigma(\mathcal{Q},
  \mathfrak{a})} i \mathcal{Q}_{\alpha}$, and hence cannot coincide
with vectors tangent to PS, which live in $\mathcal{Q}$.

\subsection{Equivalence of $PS$ and $Euclid$} \label{equiv_PS_Euclid}

Finally, we show that the integral over $PS$ equals the integral over
$Euclid$. The idea is to deform the extended PS domain into the
subspace $\mathcal{Q}_+ \oplus i [\mathfrak{p},s]$ of
$\mathcal{Q}^{\mathbb{C}}$ where $B$ is positive. Recall that
$[\mathfrak{p},s]=\mathcal{Q}_-$. We have to show that the integral
remains convergent along the path of deformation and no boundary terms
at infinity are generated. To that end we prefer to proceed in the
reverse order and deform $Euclid$ into $PS$ to the extent that this is
allowed by convergence of the integral.  Recall that
$Z_K(\mathfrak{a})$ acts trivially on $H_i \in \mathfrak{a}$ and the
matrices $E_{i}$. It also acts trivially on $[H_{i},s]$ for all $i =
1,\ldots,M$ because $k \in K$ is fixed by conjugation with $s$. For
these reasons the mapping defined as
\begin{align} \label{EPS}
  EPS_{L,c}^{\epsilon}: \, [\epsilon,1]\times&\mathfrak{a}_{L,c}^+
  \times K \times_{Z_K(\mathfrak{a})} \mathcal{Q}_+ \rightarrow
  \mathcal{Q}^{\mathbb{C}} , \notag\\ (t,H,[k; X]) \mapsto &\Ad(k)
  \Big( X_0 + \sum_{\mathclap{\alpha \in \Sigma_+(\mathcal{Q},
      \mathfrak{a})} } \big(X_{\alpha} + (1-t) T_{\alpha,c}(H)
  \phi(X_{\alpha})\big) \Big) \notag\\ &-i \Ad(k) \sum_{j \in L}
  (h^j-1) \big((1-t)E_{j} + 2 t [H_j,s] \big)
\end{align}
is well defined. By reasoning similar to that for $PS_{L,c}$ each
parametrization $EPS_{L,c}^{\epsilon}$ ($c \in C$) can be seen as an
integration cell with boundary coming only from the $[\epsilon,1]
\times\mathfrak{a}_{L,c}^+$ part.  To simplify the notation we define
\begin{align*}
  EPS^{\epsilon} := \sum_{c \in C}\sum_{L\subset I_c}
  EPS_{L,c}^{\epsilon} \, ,
\end{align*}
$EPS_{L,c} := EPS_{L,c}^{0}$ and similarly $EPS := EPS^{0}$.

\begin{lem} \label{hom_conv_lem}
The mappings $EPS_{L,c}^{\epsilon}$ have the following properties:
\begin{enumerate}[label=\roman{*})]
\item\label{partial_EPS_eps} %
The boundary of the sum $EPS^{\epsilon}$ is given by
\begin{align*}
  \partial (EPS^{\epsilon}) &= Euclid - \sum_{c \in C}\sum_{L\subset
    I_c}EPS_{L,c}(t = \epsilon) .
\end{align*}
\item \label{EPS_conv} %
Let $g(Q,A):= e^{-\Tr(Q^2)-2i\Tr(Q A)}$. Then the integrals
\begin{align*}
  \int_{EPS_{L,c}(t=\epsilon)} g(Q,A) dQ
\end{align*}
exist for $\epsilon > 0$.
\item \label{EPS_limit_zero} For each $c \in C$ and $L\subset
    I_c$ with cardinality $|L|>0$ we have
\begin{align*}
  \lim_{\epsilon \rightarrow 0} \int_{EPS_{L,c}(\epsilon)} g(Q,A) dQ =
  0 .
\end{align*}
\item \label{PS_limit}Let $PS(\epsilon):=\sum_{c\in C}
  EPS_{\emptyset,c}(\epsilon)$. The integral over $PS(\epsilon)$ in
  the limit $\epsilon \rightarrow 0$ may be computed as an integral
  over $PS$ with regularized integrand:
\begin{align*}
  \lim_{\epsilon \rightarrow 0} \int_{PS(\epsilon)} g(Q,A) dQ =
  \lim_{\epsilon\rightarrow 0} \int_{PS} g(Q,A) \chi_{\epsilon}(Q) dQ .
\end{align*}
\end{enumerate}
\end{lem}

The proof of the lemma is spelled out in the next four
subsections. Here we anticipate that once the lemma has been
established, we can do the following series of manipulations:
\begin{align*}
  \int_{Euclid} g(Q,A) dQ &= \int_{Euclid} g(Q,A) dQ -\lim_{\epsilon
    \rightarrow 0}\int_{EPS^{\epsilon}} \underbrace{d(g(Q,A) dQ)}_{=0}
  \;, \\ &\underset{i), ii)}{=}\lim_{\epsilon \rightarrow 0}
  \int_{EPS(\epsilon)} g(Q,A) dQ \\ &\underset{ii), iii),iv)}{=}
  \lim_{\epsilon \rightarrow 0} \int_{PS(\epsilon)} g(Q,A)  dQ  \\
  &\underset{iv)}{=} \lim_{\epsilon \rightarrow 0} \int_{PS} g(Q,A)
  \chi_{\epsilon}(Q) dQ ,
\end{align*}
which yields the statement of our theorem \ref{thm_2_1}.

\subsubsection{Proof of $i)$: Deformation of $PS$ into $Euclid$}

Next we prove statement $i)$ of lemma \ref{hom_conv_lem}. By an
argument similar to that in the proof of lemma \ref{Extension_lem} we
obtain
\begin{align}
  \partial (EPS^{\epsilon}) \equiv \sum_{c\in C} \sum_{L \subset
    I_c} \partial (EPS_{L,c}^\epsilon) = EPS(t=1) - EPS(t=\epsilon) .
  \label{partial_EPS}
\end{align}
We will now deal with the summand $EPS(1)$.

For $L \subsetneq I_c$ the mapping $EPS_{L,c}(t)$ degenerates in the
limit $t \to 1$. More precisely, for $i \in I_c \setminus L$ we have
$\partial_{h^i} EPS_{L,c}(1) = 0$ and thus a reduction in dimension.
Hence we have the following identity relating integration chains:
\begin{align*}
  EPS(1) = \sum_{c\in C} EPS_{I_c,c}(1) .
\end{align*}

In the following we establish the connection between $EPS(1)$ and
$Euclid$. For $c \in C$ and $H= \sum_{i \in I_c} h^i H_i \in
\mathfrak{a}_{I_c,c}^+$ we have
\begin{align*}
  EPS_{I_c, c}(1,H,[k;X]) &= \Ad(k) \Big( X - 2 i \sum_{j \in I_c}
  (h^j -1) [H_j, s] \Big) .
\end{align*}
To facilitate the interpretation of the expression on the right hand
side, we now change the left factor of the domain of definition of
$EPS_{I_c,c}(1)$ from $\mathfrak{a}_{I_c,c}^+$ to $\mathfrak{a}_c^+$.
This is done by introducing the diffeomorphism
\begin{align*}
  \psi_c : \; \mathfrak{a}_{I_c,c}^+ &\rightarrow \mathfrak{a}_c^+ \;
  , \\ H &\mapsto \sum_{i\in I_c} (h^i-1) H_{i} \, .
\end{align*}
By inserting it into the previous formula we get
\begin{align*}
  EPS_{I_c, c}(1,\psi_c^{-1}(H),[k;X]) &= \Ad(k) \big( X - 2 i[H, s]
  \big) .
\end{align*}
Note that while the composition with $\psi_c^{-1}$ does alter the
mapping $EPS_{I_c,c}(1)$, the effect is not a change of image but only
a reparametrization.

The right hand side of the expression above does not have any
explicit dependence on the simplicial cone $c$ any more. Therefore,
by recalling $\bigcup_{c \in C} \mathfrak{a}_c^+ = \mathfrak{a}^+$
and noting that the diffeomorphisms $\psi_c$ are orientation
preserving, it is clear that our mappings $EPS_{I_c, c}$ combine to a
smooth mapping
\begin{align*}
  \sum_{c \in C} EPS_{I_c,c}(1) \circ (\psi_c^{-1}, id) : \quad
  &\mathfrak{a}^+ \times K \times_{Z_K(\mathfrak{a})} \mathcal{Q}_+
  \rightarrow \mathcal{Q}^{\mathbb{C}} , \\
  &(H, [k ; X]) \mapsto \Ad(k) \big( X - 2 i [ H, s] \big) ,
\end{align*}
where $id$ stands for the identity on $K \times_{Z_K(\mathfrak{a})}
\mathcal{Q}_+$.

The final step is to undo the reparametrizations $R_{\rm II}$ and
$R_{\rm I}$ to obtain
\begin{align*}
  \Big(\sum_{c \in C} EPS_{I_c,c}(1) \circ (\psi_c^{-1}, id)\Big)
  \circ R_{\rm II}^{-1} \circ R_{\rm I}^{-1} (Y, X) = X - 2i [Y,s] ,
\end{align*}
where $X \in \mathcal{Q}_+$ and $Y \in \mathfrak{p}$. Since
$[\mathfrak{p},s] = \mathcal{Q}_-$ we conclude that $EPS(1)$
is the same as Euclid as an integration chain.

\subsubsection{Proof of $ii)$: Existence of the integral for $\epsilon >0$}
\label{proof_existence}

In this subsection we prove statement $ii)$ of lemma
\ref{hom_conv_lem}. Let us first make some general remarks and
definitions which allow a simpler discussion of the integrals to be
considered. For this purpose let $id_{L,c}$ be the identity on
$\mathfrak{a}_{L,c}^+$ and recall that $R_{\rm II}$ yields a global
factorization of the bundle $K \times_{Z_{K}(\mathfrak{a})}
\mathcal{Q}_+$ as $K/Z_{K}(\mathfrak{a})\times \mathcal{Q}_+$. Let
$d\mu([k])$ be a left invariant volume form on $K/Z_{K}
(\mathfrak{a})$ and let $dH$ and $dQ_+$ denote constant volume forms
on $\mathfrak{a}^+$ and $\mathcal{Q}_+$ respectively. Then there
exist functions $P_{L,c}$ such that
\begin{align*}
  \left(EPS_{L,c}(t=\epsilon) \circ (id_{L,c}, R_{\rm II}^{-1})
  \right)^* dQ = P_{L,c} \; dH \wedge d\mu([k]) \wedge dQ_+ \, .
\end{align*}
By inspection of $EPS_{L,c}$ and $R_{\rm II}^{-1}$ we see that
$P_{L,c}$ depends polynomially on $\epsilon$, $[k]$, the matrix
entries of $X \in \mathcal{Q}_+$, $h^i$, and on $\partial^rT_{\alpha,
  c}$, where $\partial^r$ represents any number of partial derivatives
with respect to $h^i$. For the rest of the proof it is more
convenient to switch to a formulation in terms of measures instead of
volume forms. In that respect we have
\begin{align}
  &\int_{\mathfrak{a}_{L,c}^+ \times K/Z_{K}(\mathfrak{a}) \times
    \mathcal{Q}_+} \Big(EPS_{L,c}(t=\epsilon) \circ (id_{L,c},
  R_{\rm II}^{-1}) \Big)^* (g(\cdot,A) dQ) \notag \\ =
  &\int\limits_{\mathclap{\mathfrak{a}_{L,c}^+ \times
      K/Z_{K}(\mathfrak{a}) \times \mathcal{Q}_+}}
  g\Big(EPS_{L,c}(t=\epsilon) \circ (id_{L,c}, R_{\rm II}^{-1}),A\Big) \;
  P_{L,c} \, |dH| |d\mu([k])| |dQ_+|
  ,\label{L_c_lebesgue_integral}
\end{align}
where $|dH|$ and $|dQ_+|$ are Lebesgue measures on $\mathfrak{a}^+$
and on $\mathcal{Q}_+$ and $|d\mu([k])|$ denotes Haar measure on
$K/Z_{K}(\mathfrak{a})$. To prove statement $ii)$ it is enough to
show the existence of
\begin{align}
  \int_{K/Z_{K}(\mathfrak{a})}\int_{\mathfrak{a}_{L,c}^+}
  \int_{\mathcal{Q}_+} \Big| g\Big(EPS_{L,c} (t =
  \epsilon) &\circ (id_{L,c}, R_{\rm II}^{-1}),A\Big)\Big| \\
  &\times | P_{L,c}| \, |dQ_+| |dH| |d\mu([k])|. \label{fubini_int}
\end{align}
Indeed, the Fubini-Tonelli theorem then asserts that the original
integral exists and that Fubini's theorem can be applied to
\eqref{L_c_lebesgue_integral}.

We now deal with the integral (\ref{fubini_int}). Note that replacing
$K/Z_{K}(\mathfrak{a})$ by $K$ in \eqref{fubini_int} introduces only
a constant factor which can be absorbed in the polynomial $P_{L,c}$.
The mapping $R_{\rm II}^{-1}$ extends naturally from $K/Z_{K}
(\mathfrak{a})\times \mathcal{Q}_+$ to $K\times \mathcal{Q}_+$ and,
similarly, $EPS_{L,c}(t)$ extends from $\mathfrak{a}_{L,c}^+\times K
\times_{Z_{K}(\mathfrak{a})} \mathcal{Q}_+$ to $\mathfrak{a}_{L,c}^+
\times K \times\mathcal{Q}_+$. Furthermore we can apply for $k \in K$
the transformation
\begin{align*}
  \mathcal{Q}_+ \rightarrow \mathcal{Q}_+\, , \quad
  X \mapsto k X k^{-1},
\end{align*}
in the inner integral over $\mathcal{Q}_+$. The corresponding Jacobian
is unity. Hence \eqref{fubini_int} equals
\begin{align}
  \int\limits_{K}\int\limits_{\mathfrak{a}_{L,c}^+}\int\limits_{
  \mathclap{\mathcal{Q}_+}} \Big| g\Big(EPS_{L,c}(\epsilon), A\Big)
  \Big| \; |P_{L,c}| \, |dQ_+| |dH| |d\mu(k)| , \label{fubini_int_K}
\end{align}
where $|d\mu(k)|$ is a Haar measure on $K$.

Let us now concentrate on the exponential function $g\big(EPS_{L,c}
(\epsilon), A\big)$ which is responsible for the convergence of the
integral. Referring to the second and third lines in \eqref{EPS}, we
define
\begin{align*}
  \Xi &:= \Ad(k) \Big( X_0 + \sum_{\mathclap{\alpha \in
      \Sigma_+(\mathcal{Q},\mathfrak{a})} } \Big(X_{\alpha} +
  (1-\epsilon) T_{\alpha,c}(H) \phi(X_{\alpha})\Big)\Big) ,\\
  \Upsilon &:= -i \Ad(k) \sum_{j \in L} (h^j-1)
  \Big((1-\epsilon)E_{j} + 2 \epsilon [H_j,s]\Big) ,
\end{align*}
which lets us write the integrand in the form
\begin{align}
  g\Big(EPS_{L,c}(\epsilon,H, k,X),A \Big) =
  e^{-B(\Xi+\Upsilon,\Xi+\Upsilon) - 2i B(\Xi + \Upsilon,A) }.
  \label{g_in_detail}
\end{align}
Due to $A, \Xi \in \mathcal{Q}$ and $\Upsilon \in \mathrm{i}
\mathcal{Q}$, the terms $B(\Xi,\Upsilon)$ and $iB(\Xi,A)$ are
imaginary. Therefore, they do not contribute to \eqref{fubini_int_K}.
To evaluate the remaining terms we note some useful relations.  For
$X_{\alpha} \in \mathcal{Q}_{+,\alpha}$ and $X,X' \in \bigoplus_{
\alpha \in \Sigma_+(\mathcal{Q},\mathfrak{a})}
\mathcal{Q}_{+,\alpha}$ we have
\begin{align}
  &B(X_{\alpha},X_{\beta}) = \delta_{\alpha, \beta}
  B(X_{\alpha},X_{\alpha}) ,  \label{orth_1} \\
  &B(X,X') =-B(\phi(X),\phi(X')) , \quad
  B(\phi(X),X') = 0 . \label{orth_2}
\end{align}
$B(\Xi,\Xi)$ can be re-expressed as
\begin{align}
  B(\Xi,\Xi) &\underset{\mathclap{\eqref{orth_1}}}{=} \Tr(X_0^2) +
  \sum_{\mathclap{\alpha \in \Sigma_+(\mathcal{Q}, \mathfrak{a})}}
  \Tr\big( X_{\alpha} + (1-\epsilon) T_{\alpha,c}(H) \phi(X_{\alpha})
  \big)^2 \notag\\ &\underset{\mathclap{\eqref{orth_2}}}{=} B(X_0,X_0)
  + \sum_{\mathclap{\alpha \in \Sigma_+(\mathcal{Q}, \mathfrak{a})}} B(
  X_{\alpha}, X_{\alpha}) \Big( 1- (1-\epsilon)^2T_{\alpha,c}^2(H)\Big).
  \label{B_Xi_Xi}
\end{align}
$B$ is positive on $\mathcal{Q}_+$, and for all $H \in \mathfrak{a}^+$
and $0 < \epsilon < 1$ we have
\begin{align}
  0 < \epsilon (2-\epsilon) \leq 1- (1-\epsilon)^2T_{\alpha,c}^2(H)
  \leq 1 . \label{inequality_nonzero_epsilon}
\end{align}
Thus $B(\Xi,\Xi)$ is positive definite. Since all dependence on $X \in
\mathcal{Q}_+$ occurs in $B(\Xi,\Xi)$ this guarantees the convergence of
the inner integral in \eqref{fubini_int_K} for $\epsilon >0$.

We turn to $B(\Upsilon,\Upsilon)$. Statement $iii)$ of lemma
\ref{E_lem} asserts that $B(E_i,E_j)= 0$ for $i,j \in L \subset
I_c$. Hence we have
\begin{align*}
  B(\Upsilon,\Upsilon) = B(\Upsilon_0 , \Upsilon_0) + 4 \epsilon (1 -
  \epsilon) \sum_{i,j \in L} (h^i - 1)(h^j - 1) B(- [H_j,s], E_i) ,
\end{align*}
where $\Upsilon_0 = -2i\epsilon \mathrm{Ad}(k) \sum_j (h^j-1) [H_j ,
s]$. Note that $B(\Upsilon_0,\Upsilon_0) \geq 0$ since $\Upsilon_0 \in
i\mathcal{Q}_-$. The remaining terms of $B(\Upsilon,\Upsilon)$ are
non-negative since
\begin{align*}
  B(-[H_j,s],E_i) &\underset{\mathclap{\eqref{s_decomposition}}}{=}
  B\Big(- \sum_{\mathclap{\alpha \in \Sigma_+(\mathcal{Q},
      \mathfrak{a})}}
  \alpha(H_{j}) \phi(s_{\alpha}), E_{i}\Big) \\
  &\underset{\mathclap{\eqref{E_{i,c}_explicit_decomposition}}}{=}
  \;\; -\sum_{\mathclap{\alpha,\beta \in \Sigma_+(\mathcal{Q},
      \mathfrak{a})}} e_{i}^{\beta} \sgn(\beta(H_{i})) \alpha(H_{j})
  B(\phi(s_{\alpha}),\phi(s_{\beta})) \\
  &= \;\; -\sum_{\mathclap{\alpha \in \Sigma_+(\mathcal{Q},
      \mathfrak{a})}} e_{i}^{\alpha} |\alpha(H_{j})|
  B(\phi(s_{\alpha}),\phi(s_{\alpha})) \geq 0 .
\end{align*}
Since $E_i$ is non-zero, identity
$\eqref{E_{i,c}_explicit_decomposition}$ ensures that there exists
some $\alpha \in \Sigma_+(\mathcal{Q}, \mathfrak{a})$ such that
$e_i^{\alpha} \neq 0$, $\alpha(H_i) \neq 0$ and $s_{\alpha}\neq
0$. Thus we have
\begin{align*}
  B(-[H_i,s],E_i) > 0 .
\end{align*}
Together with \eqref{inequality_nonzero_epsilon} this shows that the
second (or middle) integral in \eqref{fubini_int_K} exists. This
conclusion is not changed by the factor $e^{-2i B(\Upsilon,A})$ as
$B(\Upsilon,A)$ is linear in the variables $h^j$ whereas $B(\Upsilon
,\Upsilon)$ is quadratic. The result depends continuously on $k \in
K$ and we hence conclude that the outer integral over the compact
group $K$ exists. It follows that all integrals $\int_{EPS_{L,c}
(\epsilon)} g(\cdot,A) dQ$ exist for $\epsilon > 0$.

\subsubsection{Proof of $iii)$: The limit $\mathbf{ \lim_{\epsilon \rightarrow 0}}$}

Recall from section \ref{proof_existence} that Fubini's theorem
applies to the integral \eqref{L_c_lebesgue_integral}. We repeat the
steps which led to \eqref{fubini_int_K} except that now we do not take
the absolute value of the integrand. Thus we obtain
\begin{align}
  \int\limits_{\mathclap{EPS_{L,c}(\epsilon)}} g(Q,A) \, dQ = \int_{K}
  \int_{\mathfrak{a}_{L,c}^+} \int_{\mathcal{Q}_+} g(EPS_{L,c}(\epsilon)
  ,A) \, P_{L,c} \, |dQ_+| |dH| |d\mu(k)| . \label{fubini_iii}
\end{align}
The reason why statement $iii)$ of lemma \ref{hom_conv_lem} holds
true is a very general one: the convergence of the integral to zero
is brought about by cancelations due to an oscillatory term. More
specifically, by integrating along one special direction in
$\mathcal{Q}_+$ we obtain essentially a regularized delta
distribution. Our parametrization is well suited to exhibit this
mechanism explicitly. We will show that it is possible to perform the
limit $\epsilon \rightarrow 0$ after doing the inner Gaussian
integrations.

In the following, Einstein's summation convention is in place. We can
choose a basis of $\mathcal{Q}_+$ with coordinates $x^l$ such that
the quadratic form $Q \mapsto \Tr Q^2$ is diagonal. The choice of
basis will be made explicit below. Schematically speaking, the
Gaussian integrations over $\mathcal{Q}_+$ in \eqref{fubini_iii} are
of the form
\begin{align*}
  I_{L,c,\epsilon}(H, k) := e^{-h^i \tilde{g}_i}
  \int\limits_{\mathbb{R}^{\mathrm{dim} \, \mathcal{Q}_+}} e^{-
    (x^{l})^2f_{l} - 2i x^{l}g_{l}} P_{L,c} \prod_{l}dx^{l} ,
\end{align*}
where $\tilde{g}_i(\epsilon,k)$, $f_{l}(\epsilon, h^i,k)$ and
$g_{l}(\epsilon, h^i, k)$ are functions of $k \in K$, $\epsilon \in
[0,1]$ and $H \in \mathfrak{a}_{L,c}^+$. These functions will be
specified as we go along. Now it is possible to introduce sources
$j_l$ for $x^l$ and perform the integral:
\begin{align}
  I_{L,c,\epsilon}(H, k) &= e^{-h^i \tilde{g}_i}
  P_{L,c}'(\partial_{j_{l}}, \dots ) \int e^{- f_{l} (x^{l})^2 - 2i
    x^{l}(g_{l} + j_{l})} \prod_l dx^{l} \Big|_{j_{l =0}} \notag \\
  &=e^{-h^i \tilde{g}_i} P_{L,c}'(\partial_{j_{l}}, \dots
  )\prod_{l} \sqrt{\frac{\pi}{f_{l}}} e^{-\frac{(g_{l}
      +j_{l})^2}{f_{l}}}\Big|_{j_{l =0}} \notag\\
  &=e^{-h^i \tilde{g}_i} P_{L,c}''(\frac{1}{f_{l}}, g_{l}, \dots)
  \prod_{l}\sqrt{\frac{\pi}{f_{l}}} e^{-\frac{g_{l}^2}{f_{l}}} ,
  \label{I_evaluated}
\end{align}
where primes just indicate that these are different polynomials, and
the dots represent a dependence on $\epsilon$, $h^i$, $k$ and
$\partial^r T_{\alpha,c}$. Assuming that $|L| > 0$, we will show that
for $\epsilon \rightarrow 0$ we have $f_1\rightarrow 0$ and $g_1\neq
0$ for a suitable choice of basis of $\mathcal{Q}_+$. We also show
that $f_l \geq 0$ and $g_l \in \mathbb{R}$ for all $l$ and $\epsilon
\in [0,1]$. The exponential then dominates the polynomial and
\eqref{I_evaluated} converges to zero in the limit of $\epsilon
\rightarrow 0$. In addition, we show that $\tilde{g}_i >0$, which has
the consequence that the remaining integrals over $\mathfrak{a}_{L,
c}^+$ are convergent (since $h^i > 0$).

Thus the issue of convergence is reduced to a discussion of the
functions $\tilde{g}_i$, $f_{l}$ and $g_{l}$. We start with
$\tilde{g}_i$. Reading it off from its definition by
\begin{align*}
  2iB(\Upsilon,A)= (h^i-1) \tilde{g}_i ,
\end{align*}
we find that it has the expression
\begin{align*}
  \tilde{g}_i = 2 B(\Ad(k)((1-\epsilon)E_i + 2\epsilon[H_i,s]), A) .
\end{align*}
{}From inequality $\ref{inequality}$ of lemma \ref{E_lem} we infer
that $2 B(\mathrm{Ad}(k) E_i , A) \geq c > 0$, since $k \in K$ and
$K$ is compact. (Recall that $A$ is fixed.) We thus see that
$\tilde{g}_i > 0$ for small enough $\epsilon$.

To discuss the functions $f_l$ and $g_l$ we have to choose a basis of
$\mathcal{Q}_+$. For this purpose we fix some $j \in L$, recalling
that $L \not= \emptyset$ in the situation at hand. We then consider the
decomposition
\begin{align} \label{Q_+_basis} \mathcal{Q}_+ = \mathcal{Q}_{+,0}
  \oplus \bigoplus_{\alpha(H_j)\neq 0} \mathcal{Q}_{+,\alpha} \oplus
  \bigoplus_{\alpha(H_j)=0} \mathcal{Q}_{+, \alpha}\, .
\end{align}
We define $m := \text{dim} \oplus_{\alpha(H_j)\neq 0}
\mathcal{Q}_{+,\alpha}$ and $m':= \text{dim} \oplus_{\alpha(H_j)= 0}
\mathcal{Q}_{+,\alpha}$. Denoting by $\Pi_{\mathcal{Q}_+}$ the
orthogonal projection onto $\mathcal{Q}_+$ we introduce
 \begin{align*}
   X_1:= \Pi_{\mathcal{Q}_+}(E_j) = \sum_{\mathclap{\alpha(H_{j})
       \neq 0}} e_{j}^{\alpha} s_{\alpha} \not= 0 .
 \end{align*}
We extend $X_1$ to an orthogonal basis $\{X_l\}_{l=1,\ldots, m}$ of
$\oplus_{\alpha(H_{j}) \neq 0}\mathcal{Q}_{+,\alpha}$. We also fix an
orthogonal basis $\{X_l\}_{l=m+1, \ldots,m+m'}$ of
$\oplus_{\alpha(H_{j}) = 0} \mathcal{Q}_{+,\alpha}$ which respects
the root decomposition. For the basis vectors $\{X_l\}_{l=1, \ldots,
  m+m'}$ we have identities like those in \eqref{orth_1} and
\eqref{orth_2}.

Recall that on $\mathfrak{a}_{L,c}^+$ we have $h^j \geq 1$ and
therefore $|T_{\alpha,c}(H)| = 1$ if $\alpha(H_j) \not= 0$. As is
shown by
\begin{align}
  B(\Xi,\Xi) \underset{\eqref{B_Xi_Xi}}{=} B(X_0,X_0) &+
  (1-(1-\epsilon)^2)\sum_{\mathclap{\alpha(H_j)\neq 0}}
  B( X_{\alpha},X_{\alpha}) \notag \\
  &+ \sum_{\mathclap{\alpha(H_j)= 0}}\Big( 1- (1-\epsilon)^2
  T_{\alpha,c}^2(H)\Big) B( X_{\alpha}, X_{\alpha}) \notag\\
  = B(X_0,X_0) &+ \sum_{l=1}^{m+m'} (x^l)^2 f_l\, ,\label{read_f}
\end{align}
our choice of basis diagonalizes the quadratic form $B(\Xi,\Xi)$.
We also see that $f_l \geq 0$. In particular, for $l =1$ we have
\begin{align*}
  f_{1} =\epsilon (2 - \epsilon) B(X_1,X_1) .
\end{align*}
Note that $f_1 \to 0$ for $\epsilon \to 0$.

It is easy to check that the coefficients $g_{l}$ defined by
\begin{align} \label{read_g}
  -2iB(\Xi,A) - 2B(\Xi,\Upsilon)= - 2 i x^{l} g_{l}
\end{align}
are real. By using the statements $ii)$ and $i)$ of lemma \ref{E_lem}
we have
\begin{align*}
  \lim_{\epsilon \rightarrow 0} \; {g}_{1} = B(E_{j}, \Ad(k^{-1})A)
  > 0 .
\end{align*}
Thus we obtain the result
\begin{align}
  \lim_{\epsilon \rightarrow 0} \int_{\mathcal{Q}_+}g(EPS_{L,c}
  (\epsilon),A) \, P_{L,c} \, |dQ_+|= 0 . \label{limit_int_zero}
\end{align}
Since the dependence on $H\in \mathfrak{a}_{L,c}^+$ in
\eqref{limit_int_zero} is governed by the exponential function
$\exp(-h^i \tilde{g}_i)$ with $\tilde{g}_i > 0$ and the dependence on
$k \in K$ is continuous, the limit $\epsilon \rightarrow 0$ is
uniform. Thus, taking the limit commutes with the outer integrals and
we obtain the third part of lemma \ref{hom_conv_lem}.

\subsubsection{Proof of $iv)$: Reaching $PS$}

It remains to prove statement $\ref{PS_limit}$ of lemma
\ref{hom_conv_lem}. To that end, for any $c \in C$ we introduce
on $\mathfrak{a}_{\emptyset,c}^+ \times K$ the two functions
\begin{align}
  I_{c, \epsilon} &:= \int\limits_{\mathcal{Q}_+} g(EPS_{\emptyset,c}
  (\epsilon),A) P_{\emptyset,c} |dQ_+| , \\ I_{c,\epsilon}' &:=
  \int\limits_{\mathcal{Q}_+} g(PS_{\emptyset,c},A)\chi_{\epsilon}
  (PS_{\emptyset,c}) P_{\emptyset,c}(\epsilon=0, \cdot) d|Q_+| .
  \label{two_functions}
\end{align}
To prove the desired statement, it is sufficient to show that
\begin{align}
  \lim_{\epsilon \rightarrow 0}
  \int\limits_{K}\int\limits_{\mathfrak{a}_{\emptyset,c}^+} I_{c,
    \epsilon} |dH| |d\mu(k)|&= \lim_{\epsilon \rightarrow 0}
  \int\limits_{K} \int\limits_{\mathfrak{a}_{\emptyset,c}^+}
  I_{c,\epsilon}' |dH| |d\mu(k)| \label{reg_eq}
\end{align}
holds for all $c\in C$. We will do so by using Lebesgue's dominated
convergence theorem on both sides of \eqref{reg_eq}.

Let us first establish that the two functions defined in
\eqref{two_functions} converge pointwise to the same function in the
limit of $\epsilon \rightarrow 0$. For that, we have to distinguish
between two situations for $H \in \mathfrak{a}_{\emptyset,
  c}^+$: there either exists a non-trivial $L \subset I_c$ such that
$H \in \mathfrak{a}_{L,c}^+ \cap \mathfrak{a}_{\emptyset,c}^+$, or
there does not. In the first situation we can apply the result of the
previous section to see that $\lim_{\epsilon \rightarrow 0}
I_{c,\epsilon}(H, k) = 0$ for all $k \in K$. A similar argument yields
the same result for the function $I_{c,\epsilon}'$. In the second
situation there are no problems of convergence with the
$\mathcal{Q}_+$-integral (see \eqref{B_Xi_Xi}) and we can directly set
$\epsilon =0$, in which case the two functions coincide by
definition. Thus we always have $I_{c,0} = I_{c,0}'$.

{}From now on, for brevity, we discuss only the left hand side of
\eqref{reg_eq}, as the discussion of the right hand side is
completely analogous. Our strategy is to show that the function
$(\epsilon,H,k) \mapsto I_{c, \epsilon} (H,k)$ on the compact domain
$[0,1] \times \mathfrak{a}_{\emptyset,c}^+ \times K$ is continuous
and hence integrable. The property of continuity on a compact domain
implies that the function $I_{c, \epsilon}$ is dominated by a
constant function, which in turn is integrable as well. Thus we will
be able to draw the desired conclusion by applying Lebesgue's
dominated convergence theorem.

Following the line of reasoning of the last subsection we choose for
$\mathcal{Q}_+$ an orthogonal basis $\{X_{l}'\}_{l = 1, \ldots,
  \mathrm{dim}\mathcal{Q}_+}$ compatible with the decomposition
$\oplus_{\alpha} \mathcal{Q}_{+,\alpha}$. This means that for each
$l$ there exists a unique root $\alpha_l$ (possibly the zero root)
such that $X_l \in \mathcal{Q}_{+, \alpha_l}$. Moreover, whenever a
root $\alpha$ is such that $s$ has non-zero projection $s_{\alpha} =
\pi_{+,\alpha}(s)$, then we take $s_\alpha \in \mathcal{Q}_{+,
  \alpha}$ to be an element of our basis set $\{ X_l'\}$. We arrange
for these non-zero vectors $s_\alpha$ to be the first $m_1$ vectors of
the set $\{ X_l' \}$.

By adaptation (due to the change of basis $\{ X_l \} \to \{ X_l' \}$)
of the definitions \eqref{read_f} and \eqref{read_g} we obtain new
coefficient functions
\begin{align*}
  f_{l}' &= \left( 1- (1-\epsilon)^2 T_{\alpha_l, c}^2 \right)
  B(X_l',X_l') , \\ g_{l}' &= \Tr\left( \big( X_{l}' + (1-\epsilon)
    T_{\alpha_l, c} \, \phi(X_{l}') \big) k^{-1} A k \right) ,
\end{align*}
superseding the earlier functions $f_l$ and $g_l$. If $\alpha_l$ is
the zero root we set $T_{0,c} = 0$. Note that $g_l'$ is still real
and $f_l' \geq 0$. By recalling the dependence on $H \in
\mathfrak{a}_{\emptyset,c}^+$ of the functions $T_{\alpha,c}$ defined
in section \ref{ps_domain}, we see that if $\epsilon \rightarrow 0$
and if $h^j \rightarrow 1$ for at least one index $j \in I_c$, then
we have $f_l' \to 0$ for all $l$ with $\alpha_l (H_j)\neq 0$. In view
of this behavior, the set of problematic points where continuity of
the function $(\epsilon , H , k ) \mapsto I_{c, \epsilon}(H,k)$ is
not obvious is the set
\begin{align} \label{problematic_set}
  \{0\}\times\Big\{\sum_{i\in I_c}h^i H_i\in\mathfrak{a}_{\emptyset,
  c}^+ \mid \exists i \in I_c : h^i=1 \Big\} \times K ,
\end{align}
as will be clear presently. To be precise, the limit function
$I_{c,0}$ \emph{is not even defined} on this set. Our main work in the
rest of this subsection will be to show that it extends continuously
as zero.  We will do so by constructing a continuous function which
dominates $|I_{c,\epsilon}|$ and is zero at the problematic points.

In the following we restrict the discussion to the case of only one
summand of the polynomial in \eqref{I_evaluated}. Its modulus is
certainly smaller than
\begin{align*}
  C \, \prod_{l} \frac{\exp(-g_l'^2 / f_l')}{f_l'^{n_{l}/2}} ,
\end{align*}
with a constant $C > 0$ and natural numbers $n_l$. Here we have
dropped the factors corresponding to zero roots, as these are of no
relevance for our present purpose. Fixing some regular element
$\tilde{H} \in \mathfrak{a}_{\emptyset,c}^+$, so that $\alpha
(\tilde{H}) \neq 0$ for all $\alpha \in \Sigma_+(\mathcal{Q},
\mathfrak{a})$, we introduce the functions
\begin{align*}
  g_{l}'' :=\Tr\left( \big( X_{l}' + \sgn(\alpha_l(\tilde{H}))
    \phi(X_{l}') \big) k^{-1} A k \right) .
\end{align*}
By a short computation, these have the convenient property that
\begin{align*}
  \frac{g_{l}'^2 -g_{l}''^2}{f_{l}} =& - \left( \Tr \phi(X_{l}')
    k^{-1} A k \right)^2 - 2 \frac{\Tr(X_{l}' k^{-1} A k)\Tr\left(
      \phi(X_{l}') k^{-1} A k \right)}{\sgn(\alpha_l(\tilde{H}))
    (1+ (1-\epsilon)|T_{\alpha_l,c}|)}
\end{align*}
is a continuous function on the compact space $[0,1] \times
\mathfrak{a}_{\emptyset,c}^+ \times K$ and there exists a constant
$C' > 0$ for which we have the upper bound
\begin{align}
  C \, \prod_{l} \frac{\exp(-g_{l}'^2/f_{l}')}{f_{l}'^{n_{l}/2}} < C'
  \, \prod_{l} \frac{\exp\left(-g_{l}''^2/f_{l}' \right)}
  {f_{l}'^{n_{l}/2}} . \label{bound}
\end{align}

For the next step, we recall the constancy on $\mathfrak{a}_{
\emptyset,c}^+$ of the sign function $\sgn(\alpha(\tilde{H})) =
\sgn(\alpha_l(H_j))$ for $\alpha_l(H_j) \not= 0$. By parts $i)$ and
$ii)$ of lemma \ref{E_lem} and our choice of basis elements $X_l' =
s_{\alpha_l}$ for $l = 1, \ldots, m_1$ we then have
\begin{align}
  \sum_{l\leq m_1 \text{ and }\alpha_l(H_j)\neq 0} e_{j}^{\alpha_l}
  g_{l}''= \Tr(E_j k^{-1} A k ) > 0 \label{H_i_convergence}
\end{align}
for every $j \in I_c$. For the following discussion we let $k \in K$
be arbitrary but fixed. Inequality \eqref{H_i_convergence} guarantees
that there exists a neighborhood $U_k$ of $k$ so that for each $j \in
\{1,\ldots, \text{dim}\, \mathfrak{a}\}$ there exists an $l_j \in
\{1, \ldots, \text{dim}\mathcal{Q}_+\}$ with the property that
$g_{l_j}'' > 0$ on $U_k$. By inspection of the right hand side of
\eqref{bound} one can see that its behavior (in the limit $\epsilon
\to 0$ and close to the set of problematic points) is very similar to
that of $\exp(-1/x)/x^a$ for $x \rightarrow 0$ and positive exponent
$a$. To make this observation more tangible we now show how to
simplify the dependence of $f_l'$ on $H \in \mathfrak{a}_{\emptyset,
c}^+$. As a first step, we note that only the first factor on the
right hand side of
\begin{align*}
  f_l' = (1-(1-\epsilon) |T_{\alpha_l, c}|)(1+(1-\epsilon)
  |T_{\alpha_l, c}|) B(X_l',X_l')
\end{align*}
is relevant for the discussion of the limit behavior. Now let $H_1 ,
H_2 \in \mathfrak{a}_{\emptyset,c}^+$ and write $T_j \equiv
|T_{\alpha_l, c}(H_j)|$ and $b_l \equiv B(X_l',X_l')$ for short. By
invoking the addition formula for the hyperbolic tangent,
\begin{align*}
  \tanh(x+y) = \frac{\tanh(x) + \tanh(y)}{1 + \tanh(x) \tanh(y)} ,
\end{align*}
and observing that this formula carries over to our functions
$T_{\alpha,c}$, we obtain
\begin{align} \label{f_l_simplify} 1-(1-\epsilon) |T_{\alpha_l,
    c}(H_1+H_2)| = \frac{(1-(1-\epsilon) T_1)(1-(1-\epsilon) T_2)+
    \epsilon (2-\epsilon) T_1 T_2 }{1+T_1 T_2 } \, .
\end{align}
We claim that this identity yields the following bounds:
\begin{align}\label{eq:2bounds}
  \frac{1}{2}\prod_{i=1,2} (1-(1-\epsilon) T_i)\leq b_l^{-1} f_l'(
  \epsilon,H_1+H_2)\leq 6 \prod_{i=1,2} \sqrt{1-(1-\epsilon)T_i}\,,
\end{align}
of which the left one is immediate. To verify the right
inequality we observe that, by the identity preceding it, a stronger
statement is
\begin{align*}
  (1-(1-\epsilon) T_1)(1-(1-\epsilon) T_2)+ \epsilon (2-\epsilon)
  T_1 T_2 \leq 3 \prod_{i=1,2} \sqrt{1-(1-\epsilon) T_i} \, .
\end{align*}
Owing to $0 \leq T_i \leq 1$ this inequality is obviously true if
$\epsilon = 0$. So let $0 < \epsilon \leq 1$. Then the two square root
factors on the right hand side never vanish and we may divide by
them. Since the resulting first term on the left hand side is never
greater than one, the remaining job is to show that
\begin{align}
  \frac{\epsilon (2-\epsilon) T_1 T_2}{\sqrt{1-(1-\epsilon)
      T_1}\sqrt{1-(1-\epsilon) T_2}} \leq 2 .
\end{align}
This follows from $T_1 T_2 \leq 1$ and $\sqrt{1-(1-\epsilon)
  T_1}\sqrt{1-(1-\epsilon) T_2} \geq \epsilon$, which concludes our
proof of \eqref{eq:2bounds}. As an easy consequence of
\eqref{eq:2bounds} we have
\begin{align*}
  \frac{1}{8 b_l} f_l'(H_1) f_l'(H_2) \leq f_l'(H_1 + H_2) \leq 6
  \sqrt{f_l'(H_1) f_l'(H_2)} .
\end{align*}

We now use these bounds to simplify the dependence on $H = \sum h^i
H_i \in \mathfrak{a}_{\emptyset, c}^+$ on the right hand side of
\eqref{bound}. Iteration gives
\begin{align} \label{fixed_l_bound}
  \frac{\exp\left(-g_{l}''^2/f_{l}'\right)}{f_{l}'^{n_{l}/2}} \leq
  \frac{\exp(-\tilde{C} g_l''^2 / \prod_{i\in I_c}(1-(1-\epsilon)
    |T_{\alpha_l,c}(h^i H_i)|)^{n_{i,c}})}{\prod_{i\in
      I_c}(1-(1-\epsilon) |T_{\alpha_l,c}(h^i H_i)|)^{n_{i,c}'}} .
\end{align}
On the right hand side $h^i \alpha_l(H_i)$ is meant without summation
convention and $\tilde{C},n_{i,c}$ and $n_{i,c}'$ are positive
constants. By \eqref{def_E} and \eqref{E_{i,c}_explicit_decomposition}
it follows that $|\alpha_{l_j}(H_j)| = \text{max}_{\alpha}
|\alpha(H_j)|$. This property can be used to see that
\eqref{fixed_l_bound} is bounded by
\begin{align*}
  \frac{\exp(-\tilde{C} g_l''^2 / \prod_{i\in I_c}(1-(1-\epsilon)
    |T_{\alpha_l,c}(h^i H_i)|)^{n_{i,c}})}{\prod_{i\in
      I_c}(1-(1-\epsilon) |T_{\alpha_{l_i},c}(h^i H_i)|)^{n_{i,c}'}} .
\end{align*}
The exponential part of the right hand side is continuous and hence
we have the bound
\begin{align*}
  C \, \prod_{l=1}^{\text{dim}\mathcal{Q}_+}
  \frac{\exp(-g_{l}'^2/f_{l}')}{f_{l}'^{n_{l}/2}} < \tilde{C}'
  \prod_{i=1}^{\text{dim}\mathfrak{a}} \frac{\exp(-\tilde{C}
    g_{l_i}''^2 / (1-(1-\epsilon) |T_{\alpha_{l_i},c}(h^i
    H_i)|)^{n_{i,c}})}{(1-(1-\epsilon) |T_{\alpha_{l_i},c}(h^i
    H_i)|)^{m_{i,c}'}}
\end{align*}
where $\tilde{C}'$ and $m_{i,c}'$ are positive constants. Now it is
easy to see that the right hand side is essentially a product of
continuous functions of the form $\exp(-c/x^a)/x^{b}$ (with $a, b, c
> 0$) which are composed with continuous functions of the form
$(1-(1-\epsilon) | T_{\alpha_{l_i},c}(h^i H_i)|)$. Thus we obtain a
dominating function for $I_{c,\epsilon}$ on a neighborhood of $k$. In
particular this yields continuity of $I_{c,\epsilon}$ in each point
of the set $\{0\}\times \{\sum_{i \in I_c} h^i H_i\in \mathfrak{a}_{
\emptyset, c}^+ \mid \exists i \in I_c : h^i = 1 \}\times \{k\}$.
Since $k$ was taken to be arbitrary we obtain that $I_{c,\epsilon}$
is a continuous function on $[0,1] \times \mathfrak{a}_{\emptyset,
c}^+ \times K$. Thus $I_{c, \epsilon}$ attains a maximum. The maximum
is a dominating function and hence Lebesgue's dominated convergence
theorem can be applied. This finishes the proof of statement $iv)$ in
lemma \ref{hom_conv_lem}, which was the last step needed to complete
the proof of Theorem \ref{thm_2_1} .

\begin{rem}
  To obtain the theorem when $\mathfrak{g}=\mathfrak{k}\oplus
  \mathfrak{p}$ is the direct sum of an Abelian and a semisimple Lie
  algebra, let $\mathfrak{a}'\oplus \mathfrak{a}$ denote the
  corresponding decomposition of a maximal Abelian subalgebra of
  $\mathfrak{p}$ and replace $\mathfrak{a}^+$ by $\mathfrak{a}' \times
  \mathfrak{a}^+$ and $H$ by $H' + H$ everywhere in the proof. In
  addition let $\tilde{\mathfrak{k}}$ denote the semisimple part of
  $\mathfrak{k}$ and replace $\mathfrak{k}$ by $\tilde{\mathfrak{k}}$
  everywhere in the proof.
\end{rem}
\begin{rem}
  It is possible to choose different regularization functions
  $\chi_{\epsilon}$. The choice made here seems natural, as it has the
  highest invariance possible and was also used in earlier work.
\end{rem}
\begin{rem}
  The convergence properties can be seen quite clearly in the
  discussion of $I_{c,\epsilon}(\epsilon, H, [k])$. The convergence is
  not uniform in $A$. To have uniform convergence, we need $As \geq
  \delta > 0$. In applications with $A s \geq 0$, one has to replace
  $A$ by $A+\delta s$. For fixed $\delta > 0$ this gives uniform
  convergence in $A$.
\end{rem}

\subsection{Different representations of the integral, and alternating
  signs}\label{sect:4.4}

In this section we establish two different representations of the
integral over the PS domain. These are stated in corollaries
\ref{cor_2_1} and \ref{cor_2_2}.

Recall that the `Jacobian' $J'(\lambda)$ which appears in both
representations may have alternating sign. In the following proof of
corollary \ref{cor_2_1} we pinpoint the origin of these surprising
signs. First we review the setting. Recall that the elements of
$\mathfrak{k}$ are antihermitian and those of $\mathcal{Q}_+$
hermitian. Since $\tilde{\mathfrak{g}} = \mathfrak{k} \oplus
\mathcal{Q}_+$ is closed under hermitian conjugation, it is the
direct sum of an Abelian and a semisimple Lie algebra. The semisimple
part of $\tilde{\mathfrak{g}}$ is denoted by
$\tilde{\mathfrak{g}}_s$. We choose a maximal Abelian subalgebra
$\mathfrak{h}$ of $\mathcal{Q}_+$ containing $s$. The decomposition
of $\tilde{\mathfrak{g}}$ into an Abelian and a semisimple part
induces a decomposition of $\mathfrak{h} = \mathfrak{h}_a \oplus
\mathfrak{h}_s$ and $\mathfrak{k} = \mathfrak{k}_a \oplus
\mathfrak{k}_s$. Here $\mathfrak{h}_a$ and $\mathfrak{k}_a$ lie in
the Abelian part of $\tilde{\mathfrak{g}}$ while $\mathfrak{h}_s$ and
$\mathfrak{k}_s$ lie in the semisimple part of $\tilde{\mathfrak{g}}
$. Let $(\mathfrak{h}_s^+)^o$ denote a positive open Weyl chamber in
$\mathfrak{h}_s$ with respect to the semisimple Lie algebra
$\tilde{\mathfrak{g}}_s$. We also define $K_s =
\exp{\mathfrak{k}_s}$. Then we have the following reparametrization:
\begin{align*}
  \tilde{R}:\; \mathfrak{p} \times K_s /Z_{K_s}(\mathfrak{h}_s)
  \times (\mathfrak{h}_s^+)^o \times \mathfrak{h}_a &\rightarrow
  \mathfrak{p} \oplus \mathcal{Q}_+ \, ,\\
  (Y,[k],H_s,H_a) &\mapsto (Y, k (H_s +H_a)k^{-1}) .
\end{align*}
Recall that $K = \exp\mathfrak{k}$ is closed by assumption and $G$
denotes the closed and analytic subgroup of $GL(n,\mathbb{C})$ with
Lie algebra $\mathfrak{g} = \mathfrak{k} \oplus \mathfrak{p}$. The
subgroup $K_a = \exp\mathfrak{k}_a \subset K$ is central and closed.
By the diffeomorphism $\mathfrak{p} \to \exp{\mathfrak{p}}$ and the
isomorphism $K_s / Z_{K_s} (\mathfrak{h}_s) \cong K / (K_a Z_{K_s}
(\mathfrak{h}_s))$ we have the reparametrization
\begin{align*}
  R : \; \exp(\mathfrak{p}) K/(K_a Z_{K_s}(\mathfrak{h}_s)) \times
  (\mathfrak{h}_s^+)^o \times \mathfrak{h}_a &\rightarrow \mathfrak{p}
  \oplus \mathcal{Q}_+ \,, \\ (e^{Y}[k], H_s, H_a) &\mapsto(Y, k (H_s
  + H_a) k^{-1}) .
\end{align*}
By $K_a Z_{K_s}(\mathfrak{h}_s) = Z_K(\mathfrak{h}_s)$ and the Cartan
decomposition $G = \exp( \mathfrak{p})K$ (see \cite{knapp05}) we
obtain yet another para\-metrization of the PS domain,
\begin{align*}
  PS \circ R : \; G/Z_{K}(\mathfrak{h}_s) \times (\mathfrak{h}_s^+)^o
  \times \mathfrak{h}_a& \rightarrow \mathcal{Q} ,\\
  ([g], H_s, H_a) &\mapsto g (H_s + H_a) g^{-1} ,
\end{align*}
which is the one most frequently used in the literature.

To proceed with the proof of corollary \ref{cor_2_1}, we have to
diagonalize the commutator action of $\mathfrak{h}$ on
$\tilde{\mathfrak{g}}$ and on $\mathfrak{p}\oplus \mathcal{Q}_-$. For
this purpose we note that the values $\alpha(H_s + H_a)$ of the roots
$\alpha$ are real since $[H_s + H_a, \cdot]$ is hermitian with
respect to the hermitian form $\Tr (X^\dagger Y)$. Moreover
$\alpha(H_a) = 0$ for $\alpha \in \Sigma_+(\mathfrak{k} \oplus
\mathcal{Q}_+,\mathfrak{h})$.

The pullback of $dQ$ by $PS \circ R$ is then
\begin{align*}
 (PS &\circ R)^* dQ = \Delta(H_s + H_a) \, d\mu([g]) \wedge dH ,
\end{align*}
where $d\mu([g])$ is a left invariant volume form on $G /
Z_K(\mathfrak{h}_s)$ and $dH$ is a constant volume form on
$\mathfrak{h}$. Denoting by $d_{\alpha}$ the dimension of the root
space corresponding to $\alpha$, we get
\begin{align*}
    \Delta(H_s + H_a) = \prod_{\alpha \in \Sigma_+(\mathfrak{k}
    \oplus \mathcal{Q}_+,\mathfrak{h})} \alpha(H_s)^{d_{\alpha}}
    \prod_{\beta \in \Sigma_+(\mathfrak{p}\oplus
    \mathcal{Q}_-, \mathfrak{h})} \beta(H_s + H_a)^{d_{\beta}} .
\end{align*}
Note that $\Delta$ differs from $J'$ in corollary \ref{cor_2_1} only
by taking the modulus of the roots in $\Sigma_+(\mathfrak{k}\oplus
\mathcal{Q}_+,\mathfrak{h})$. But these roots $\alpha \in
\Sigma_+(\mathfrak{k}\oplus \mathcal{Q}_+,\mathfrak{h})$ are positive
when evaluated on $(\mathfrak{h}_s^+)^o$. Therefore we have the
following equality:
\begin{align*}
  \int_{PS \circ R} f(Q)\, dQ = \int_{id} f(g(H_s + H_a)g^{-1})
  J'(H_s+H_a)\, d\mu([g]) \wedge dH ,
\end{align*}
where $id$ denotes the identity on $G / Z_K(\mathfrak{h}_s) \times
(\mathfrak{h}_s^+)^o\times \mathfrak{h}_a$.

At this point a crucial difference between the roots in $\Sigma_+
(\mathfrak{k} \oplus \mathcal{Q}_+,\mathfrak{h})$ and those in
$\Sigma_+(\mathfrak{p} \oplus \mathcal{Q}_-, \mathfrak{h})$ is
detected: since the definition of the Weyl chamber $\mathfrak{h}_+^o$
refers only to the former roots, it is possible for the latter roots
to change sign on $\mathfrak{h}_+^o$. These sign changes are
particulary evident in our approach as we are integrating
differential forms instead of densities (or measures).

Now it is convenient to replace the volume form $d\mu([g])$ by the
left invariant measure $|d\mu([g])|$ and $dH$ by Lebesgue measure
$|dH|$ on $\mathfrak{h}:$
\begin{align*}
    \int\limits_{\mathclap{PS \circ R}} f(Q)\, dQ = \quad
    \int\limits_{\mathclap{G / Z_K(\mathfrak{h}_s) \times
    (\mathfrak{h}_s^+)^o\times \mathfrak{h}_a}} f(g(H_s +
    H_a) g^{-1}) J'(H_s + H_a) |d\mu([g])| |dH | .
\end{align*}
The replacement of $G/(K_a Z_{K_s}(\mathfrak{h}_s))$ by $G$ simply
leads to a change of normalization constant $c' \in \mathbb{R}
\setminus \{0\}$:
\begin{align*}
  \int\limits_{\mathclap{PS \circ R}} f(Q)\, dQ = c'
  \int\limits_{\mathclap{G \times (\mathfrak{h}_s^+)^o\times
      \mathfrak{h}_a}} f(g(H_s + H_a)g^{-1}) J'(H_s+H_a)
  |d\mu(g)| |dH | ,
\end{align*}
where $|d\mu(g)|$ denotes Haar measure on $G$.

Let $N_{K_s}(\mathfrak{h}_s)$ denote the normalizer of $\exp
\mathfrak{h}_s$ in $K_s$. In the following we make use of the Weyl
group $N_{K_s}(\mathfrak{h}_s)/Z_{K_s}(\mathfrak{h}_s)$. This Weyl
group acts on $\mathfrak{h}_s$ and generates $\mathfrak{h}_s$ from
$\mathfrak{h}_s^+$. To exploit this property we need that $J'$ is
invariant under the action of the Weyl group. Recall that $J'$ is
given by
\begin{align*}
  J'(H) = \prod_{\alpha \in \Sigma_+(\mathfrak{k}\oplus
    \mathcal{Q}_+,\mathfrak{h})} |\alpha(H)^{d_{\alpha}}|
    \prod_{\beta \in \Sigma_+(\mathfrak{p}\oplus
    \mathcal{Q}_-, \mathfrak{h})} \beta(H)^{d_{\beta}} .
\end{align*}
The first factor is trivially invariant, whereas for the second
factor an additional argument is needed. For that purpose we define a
root $\beta \in \Sigma(\mathfrak{p}\oplus \mathcal{Q}_-,
\mathfrak{h})$ to be positive if $\beta(s) > 0$. This definition
makes sense because $\beta(s) \neq 0$ for all roots $\beta \in
\Sigma( \mathfrak{p}\oplus \mathcal{Q}_-, \mathfrak{h})$. Since $s$
is $\Ad(K)$ invariant we conclude that the action of the Weyl group
does no more than permute the roots in $\Sigma_+ (\mathfrak{p} \oplus
\mathcal{Q}_-, \mathfrak{h})$. Hence $\Sigma_+(\mathfrak{p} \oplus
\mathcal{Q}_-, \mathfrak{h})$ is Weyl-invariant and so is $J'$. Now
the Haar measure $|d\mu(g)|$ is $G$-bi-invariant and hence
Weyl-invariant. Therefore, introducing another normalization constant
$c''\in \mathbb{R}\setminus\{0\}$ we have
\begin{align*}
  \int\limits_{\mathclap{PS \circ R}} f(Q)\, dQ = c''
  \int\limits_{\mathclap{G \times \mathfrak{h}}} f(g H g^{-1})
  J'(H) |d\mu(g)| |dH | .
\end{align*}
By setting $ f = g \cdot \chi_{\epsilon}$ we obtain corollary
\ref{cor_2_1}.

Since $PS$ is nearly everywhere injective and regular by assumption,
so is $PS \circ R$. Application of the change of variable theorem
then yields corollary \ref{cor_2_2}.

\begin{appendix}
\section{Contributions from $\partial \mathfrak{a}^+$}
\label{app_contributions}

Here we give the detailed argument showing that for our purpose of
integrating over $PS$ and $EPS^\epsilon$ the contributions from the
boundary $\partial \mathfrak{a}^+$ of the Weyl chamber
$\mathfrak{a}^+$ are irrelevant, as they are of codimension at least
two.

Without loss, we fix any $c \in C$ and let $H_{i} \in
\mathfrak{a}^+_{c} \cap \partial\mathfrak{a}^+$ be any one of the
generators of $\mathfrak{a}_c^+$ which also lie in $\partial
\mathfrak{a}^+$ (if there is no such generator then there is nothing
to prove). By removing this generator we get a boundary component
\begin{align*}
    \mathfrak{a}_{i,c}^+ := \mathfrak{a}_{i,c} \cap \mathfrak{a}^+
    \subset \partial\mathfrak{a}_c^+ \cap \partial\mathfrak{a}^+ \;,
    \quad \mathfrak{a}_{i,c} := \left\{\sum\nolimits_{j \in I_c
    \setminus \{i\}} h^j H_{j} \mid h^j \in \mathbb{R} \right\}.
\end{align*}

Next recall the definition of $PS_c$ in \eqref{PS_c_parametrisation}.
We now show that by restricting to $\mathfrak{a}_{i,c}^+$ in the
leftmost factor of the domain of definition of $PS_c$ we get a domain
of codimension at least two. Here the main observation is that the
dimension of the isotropy group of $\mathfrak{a}$ changes at the
boundary of $\mathfrak{a}^+$ and, in particular, $\dim Z_{K}
(\mathfrak{a}_{i,c}) > \dim Z_{K}(\mathfrak{a})$. This is seen as
follows. Each face of $\partial \mathfrak {a}^+$ lies in the zero
locus $\mathrm{ker}\, \alpha$ of some root $\alpha \in \Sigma(
\mathfrak{g}, \mathfrak{a})$ and we can arrange for $\mathfrak{a}_{
i,c} \subset \mathrm{ker}\, \alpha$. If $\mathfrak{g}_\alpha$ is the
root space of $\alpha$, the group generated by $\Fix_{\theta}
(\mathfrak{g}_\alpha \oplus \mathfrak{g}_{-\alpha}) \not\subset
\mathrm{Lie} Z_K(\mathfrak{a})$ leaves the face $\mathfrak{a}_{i,
c}^+$ invariant. When restricting $PS \circ R_{\rm I}$ in the first
factor to $\mathfrak{a}_{i,c}^+$ we may replace the second factor $K/
Z_{K} (\mathfrak{a})$ by the lower dimensional space $K/ Z_{K}
(\mathfrak{a}_{i,c})$ without changing the image of the
parametrization. Thus the reduction $\mathrm{dim}\, \mathfrak{a}_{
i,c}^+ < \mathrm{dim}\, \mathfrak{a}$ is accompanied by a reduction
of dimension of the $K$-orbits on $H \in \mathfrak{a}_{i, c}^+$.
Altogether, the dimension is reduced by no less than two. Moreover,
the eigenspace decomposition of $\mathcal{Q}$ with respect to
$\mathfrak{a}$ is a refinement of the eigenspace decomposition
w.r.t.\ the smaller abelian algebra $\mathfrak{a}_{i,c}$. Hence our
further reparametrizations of PS (by $R_{\rm III}$ and $R_{\rm IV}$,
which rely on an eigenspace decomposition of $\mathcal{Q}$) are
compatible with the restriction of $\mathfrak{a}$ to $\mathfrak{a}_{
i,c}$. This completes the argument for $PS_c$.

Turning to $EPS^{\epsilon}$, we have to argue that the analogous
restriction is still well defined. For that, it is enough to note
that for $X \in Z_K(\mathfrak{a}_{i,c})$ we have $[X , E_j] = 0$ if
$j\neq i$. By this token we see that also for $EPS^{\epsilon}$ the
contributions from $\partial \mathfrak{a}^+$ are of codimension at
least two.

\section{Equivalence of $SW$ and $Euclid$} \label{app_SW_Euclid}

A detailed discussion of the SW domain and the validity of the
pertinent Hubbard-Stratonovich transformation can be found in
\cite{zirn96}. Here we give another proof by deforming $SW$ into
$Euclid$. By using some of the constructions of the proof for the PS
domain, this deformation can be stated very explicitly.

We start with a brief discussion of the convergence of the Gaussian
integral (\ref{motiv_identity}) over
\begin{align*}
  SW :\; \mathfrak{p} \oplus \mathcal{Q}_+ &\rightarrow
  \mathcal{Q}^{\mathbb{C}} ,\\ (Y,X) &\mapsto X - i b e^Y s e^{-Y} .
\end{align*}
For $X \in \mathcal{Q}_+$ and $Y \in \mathfrak{p}$ we have that
$B(X,X) \geq 0$ and
\begin{align*}
  B(i b e^Y s e^{-Y},i b e^Y s e^{-Y}) = -b^2 B(s,s)
\end{align*}
is constant. The cross term $B(X,i b e^Y s e^{-Y})$ is purely
imaginary, and
\begin{align*}
  -iB(-i b e^Y s e^{-Y},A) = - b\Tr(e^{-2Y}As) < 0
\end{align*}
for $b > 0$ yields convergence in the $\mathfrak{p}$ directions.

To see the properties of $SW$ more explicitly, we use the
reparametrization $R_{\rm I}$ and the decomposition of $s$ to obtain
\begin{align*}
  &SW \circ R_{\rm I}: \; \mathfrak{a}^+ \times K/Z_{K}(\mathfrak{a})
  \times Q_{+}  \rightarrow \mathcal{Q}^{\mathbb{C}} , \\ &(H,[k],X)
  \mapsto X - i b\Ad(k) \Big( s_0 + \sum_{\mathclap{\alpha \in
  \Sigma_+(\mathcal{Q}, \mathfrak{a})}} \big(\cosh(\alpha(H)) s_{\alpha}
  + \sinh(\alpha(H)) \phi(s_{\alpha}) \big) \Big) .
\end{align*}
From this parametrization we see that the image of the boundary of
$\mathfrak{a}^+$ is again of codimension at least two, which clearly
shows that $\partial (SW) = 0$.

A homotopy from $SW$ to $Euclid$ is given by
\begin{align*}
  ESW :\;[0,1] &\times \mathfrak{a}^+ \times K/Z_{K}(\mathfrak{a})
  \times \mathcal{Q}_+  \rightarrow \mathcal{Q}^{\mathbb{C}} ,\\
  (t,H,[k],X) &\mapsto X - ib\Ad(k) \Big[ (1-t)s_0 + \\
  \sum_{\mathclap{\alpha \in \Sigma_+(\mathcal{Q}, \mathfrak{a})}}
  &[(\cosh((1-t)\alpha(H))-t) s_{\alpha} + \frac{\sinh((1-t)
    \alpha(H))}{1-t} \, \phi(s_{\alpha})]\, \Big] .
\end{align*}
Note that $ESW(t=0) = SW$ and $ESW(t=1,H,[k],X) = X - i b[kHk^{-1},
s]$. Since $[\mathfrak{p},s] = \mathcal{Q}_-$ we obtain $ESW(1) =
Euclid$.

To complete the argument we show that the integral over $ESW$ is
convergent. For this we note that for $Q_t = ESW(t,H,[k],X)$ we have
\begin{align*}
    B(Q_t,Q_t) = B(X,X) + \sum_{\mathclap{\alpha \in \Sigma_+
    (\mathcal{Q}, \mathfrak{a})}}(2t -t^2) \frac{\sinh^2((1-t)
    \alpha(H))}{(1-t)^2} B(s_{\alpha},s_{\alpha}) + \dots \;,
\end{align*}
where the dots represent unimportant terms; these are terms which are
purely imaginary, terms which are linear in $\sinh$ and all terms
containing $s_0$. Owing to $B(X,X) \geq 0$ and $B(s_\alpha,s_\alpha)
> 0$ we obtain convergence for $t > 0$. For $t = 0$ convergence is
ensured by the $B(Q,A)$ term, as was discussed above for the $SW$
parametrization.

\medskip\noindent\textbf{Acknowledgment.} This research was
financially supported by a grant from the Deutsche
Forschungsgemeinschaft (SFB/TR 12). J.M.H.\ gratefully acknow\-ledges
useful discussions with P.\ Heinzner.
\end{appendix}

\end{document}